\documentclass[10pt]{article}
\usepackage[utf8]{inputenc}
\usepackage[T1]{fontenc}
\usepackage{amsmath}
\usepackage{amsfonts}
\usepackage{amssymb}
\usepackage[version=4]{mhchem}
\usepackage{hyperref}
\hypersetup{colorlinks=true, linkcolor=blue, filecolor=magenta, urlcolor=cyan,}
\usepackage{graphicx}
\usepackage[export]{adjustbox}
\graphicspath{ {./images/} }

\usepackage{enumitem}
\usepackage{placeins} 
\usepackage[english]{babel}
\usepackage{microtype}
\usepackage{url} 
\usepackage{amsthm}

\usepackage[numbers]{natbib}
\usepackage{setspace}
\newtheoremstyle{nodotstyle} 
  {12pt} 
  {12pt} 
  {} 
  {} 
  {} 
  {} 
  { } 
  {} 

\usepackage{setspace}
\usepackage{float}
\usepackage{subcaption}

\theoremstyle{nodotstyle}
\usepackage{titlesec}
\usepackage{braket}

\titleformat{\subsection}[hang]{\bfseries}{\thesubsection}{1em}{}
\titlespacing*{\subsection}{0pt}{*1.5}{*1}

\begin{document}

\title{An Indeterminacy-based Ontology for Quantum Theory}

\author{Francisco Pipa\thanks{franciscosapipa@gmail.com}\\
\vspace{0.3cm}\\ Department of Philosophy, University of Kansas}
\date{}

\maketitle

\begin{abstract}
\noindent I present and defend a new ontology for quantum theories (or
``interpretations'' of quantum theory) called Generative Quantum Theory
(GQT). GQT postulates different sets of features, and the combination of
these different features can help generate different quantum theories.
Furthermore, this ontology makes quantum indeterminacy and determinacy
play an important explanatory role in accounting for when quantum
systems whose values of their properties are indeterminate become
determinate. The process via which determinate values arise varies
between the different quantum theories. Moreover, quantum states
represent quantum properties and structures that give rise to
determinacy, and each quantum theory specifies a structure with certain
features. I will focus on the following quantum theories: GRW, the
Many-Worlds Interpretation, single-world relationalist theories such as
Relational Quantum Mechanics, Bohmian Mechanics, hybrid
classical-quantum theories, and Environmental Determinacy-based (EnD)
Quantum Theory. I will argue that GQT should be taken seriously because
it provides a series of important benefits that current widely discussed
ontologies lack, namely, wave function realism and primitive ontology,
without some of their costs. For instance, it helps generate quantum
theories that are compatible with relativistic causality, such
as EnD Quantum Theory. Also, GQT has the benefit of providing new ways
to compare and evaluate quantum theories, which may lead to
philosophical and scientific progress.
\end{abstract}

\section{Introduction}\label{introduction}

What exists at the fundamental level according to our best scientific
theories? Or, more concretely, what is the right ontology behind the
puzzling phenomena represented by quantum theory (QT), arguably our most
widely applicable fundamental theory? It is unclear how to understand and
answer satisfactorily these questions. There are several interpretations
of QT, or more accurately, different quantum theories (QTs). Also, there
are longstanding foundational and philosophical issues surrounding the
elements of the theory, such as the wave function or quantum state. To
address these questions, there are diverse, what I will call ontological
frameworks that provide clear ontologies applicable to various quantum
theories.


A widely debated framework is Wave function Realism,\footnote{See, e.g., \cite{Albert1996, Ney2021ThePhysics, Albert2023AMechanics}.} which considers that what fundamentally
exists is a wave function living in a large multidimensional
configuration space.\footnote{The wave function that represents this field
  mathematically concerns quantum states expressed in a basis.
  Disregarding the spin, in non-relativistic QT, the wave function is
  typically considered a \emph{square-integrable} and smooth function
  whose domain is the \(\mathbb{R}^{3N}\) configuration space where N is
  the number of particles, and the range is complex numbers. More concretely, in the non-relativistic case, the wave function in the configuration space is obtained by projecting the quantum state onto the position basis of the particles under analysis. For a single particle, the wave function $\psi(\mathbf{r}, t)$ is given by $\Psi(\mathbf{r}, t) = \braket{\langle \mathbf{r} | \Psi(t)} \rangle$. This projection gives the probability amplitude for finding the particle at position $\mathbf{r}$ at time $t$. For a system of $N$ particles, the wave function $\Psi(\mathbf{r}_1, \mathbf{r}_2, \ldots, \mathbf{r}_N, t)$ is obtained by projecting the quantum state onto the combined position basis of all particles:  $ \Psi(\mathbf{r}_1, \mathbf{r}_2, \ldots, \mathbf{r}_N, t) = \langle \mathbf{r}_1, \mathbf{r}_2, \ldots, \mathbf{r}_N | \Psi(t) \rangle$. $\mathbf{r}_i$ represents the position of the $i$-th particle. Thus, in this non-relativistic case, for the wave function realist, if we have a universe with just one particle, we consider that that universe just has 3 dimensions. If we have a universe with N particles, for the wave function realist, the space of that universe has literally 3N dimensions, each dimension corresponding to one possible configuration of the particles. Even if particles are far apart in the three-dimensional space, such as in the Bell scenario, they are near each other in the configuration space. This ``locality" in the configuration space makes this view attractive to some. The wave function is a field because it assigns complex amplitudes to the regions of this multidimensional space.} Another widely debated framework called primitive ontology,\footnote{See, e.g., \cite{Allori2013PrimitiveTheories, Durr1992QuantumUncertainty, Goldstein201391RealityTheory}.} typically considers that quantum
states/wave functions/density operators have a nomological character.
Moreover, the primitive ontology that these objects describe/govern
concerns entities with determinate features that live in a determinate
location of the three-dimensional space. Other alternatives consider
that the density operator is a property of spacetime points,\footnote{\cite{Wallace2010QuantumRealism}. See also, e.g., \cite{sep-qt-issues} and references
  therein.} etc. Since we currently don't know what the right QT is, a
plausible strategy to investigate its ontology is to formulate and
analyze ontological frameworks, which, given their generality and
clarity, will likely provide that information.

What the current major ontological frameworks have in common is that
they consider that there are determinate properties or features or laws
or fields (e.g., wave function, primitive ontology, etc.) that are
fundamental and play a key explanatory role. Indeterminate properties,
in a sense that will be clarified, arise from them and have a secondary
explanatory role. For instance, according to wave function realism,
indeterminate properties arise from a multidimensional field, and what
plays a key explanatory role is this field.\footnote{See also \cite{Glick2017AgainstIndeterminacy}.} However, historically, the so-called Eigenstate-Eigenvalue
Link (EEL) played an important role in interpreting QT, especially
within the more ``orthodox'' interpretations.\footnote{See \cite{2016SHPMP..55...92G}
  for a historical overview of the importance of this link.} According
to this link:\\

\emph{A system $S$ has a determinate value q of an observable O if and
only if the quantum state of $S$ is in an eigenstate of} \(O\) \emph{with an
eigenvalue} \(q\)\emph{.}\\

This link often leads to the assumption that if the quantum state of S
is in a quantum state that is not an eigenstate of some observable, the
system has, in a sense, an \emph{indeterminate value} of that
observable. It also has led to the view that QT presents a new kind of
indeterminacy, an ontological indeterminacy.\footnote{See, e.g., \cite{Barnes2011AIndeterminacy}, \cite{Lewis2016QuantumMechanics}, and \cite{Calosi2019QuantumIndeterminacy}.} The
source of this indeterminacy is not in our knowledge or the semantics of
our language\footnote{See, e.g., \cite{Fine1975VaguenessLogic} and Williamson (1994).} but
in the world itself. However, despite this link's importance, an
ontological framework applicable to the major QTs, and
where quantum indeterminacy plays an important explanatory role, hasn't
been proposed and defended.

I propose and defend an alternative ontological framework called
generative quantum theory (GQT). In this framework, quantum
indeterminacy plays an important explanatory role rather than the
opposite, reversing the arrow of explanation from what is typically
proposed by the ontological frameworks of QT. I will compare GQT to the
most widely discussed ontological frameworks, wave function realism and
primitive ontology, and explain that it doesn't suffer from some of
their notorious costs while proving some important benefits.

The rough idea of GQT is that the world is constituted by default by
entities with so-called \emph{indeterminate (value) properties}, which
give rise to entities with \emph{determinate (value) properties}, which are considered to be objective features of the world. GQT
will also offer the possibility of giving rise to QTs where
entities with determinate value properties also exist by default, along
with entities with indeterminate value properties.\footnote{Such as
  Bohmian mechanics, as we shall see.} Via a relation between
determinate and indeterminate value properties and other features of
systems, I will also provide a new analysis of quantum indeterminacy and
determinacy. Furthermore, contrary to the previous ontological
frameworks, this framework postulates different new sets of features. In
this article, I will propose seven different but interrelated features.
The combination of these different sets can help generate different
QTs, which, as I will argue, will provide several benefits.

Often associated with Wave function Realism and Primitive Ontology is a
reificatory view of the wave function or a literalist reading of it
either as a law or an object. GQT will adopt a different view of quantum
states/wave functions and density operators,\footnote{Note that I will
  often refer to density operators as quantum states.} assigning them
various roles. They will allow for inferences about and representation
of the different possibilities of determinate values arising. Also, they
help represent the so-called \emph{quantum properties} of systems that
are related to value properties, ``giving rise to them.'' However, this
representation won't be a literalist or \emph{self-standing} one.
Quantum states have the support of values, observables and other tools,
such as directed graphs, DAGs (directed acyclic graphs, i.e., directed
graphs with no cycles), and Quantum Causal Models, to make inferences
and represent the various features mentioned above. Importantly, and
this is a key innovation, these tools will support inferences and
representations about structures of interactions that give rise to
determinacy.\footnote{See Section 2.4.} Different QTs will appeal to
different structures of this kind. Given their epistemic role, quantum
states of a system won't collapse in a physical sense during
interactions. There is instead a state update of the original state of a
system that can be implemented, for example, upon its decoherence,
specific interactions, under collapse, branching, etc. I will argue that
this view on the nature of quantum states is less problematic than the
one adopted by WR and PO.




I will start by presenting the basics of GQT via the
Ghirardi-Rimini-Weber (GRW) theory (Section 2.1). Then, in the rest of
section 3, I will present the Many-Worlds Interpretation (MWI),
relationalist single-world, and Bohmian Mechanics versions of
GQT.\footnote{See, e.g., \cite{Wallace2012TheInterpretation, Goldstein2021BohmianMechanics, sep-qm-collapse} and references therein.} Finally (Section 2.4), I will
show how GQT helps moving beyond the standard interpretations by helping
generating Environmental Determinacy-based Quantum
Theory (EnDQT). I have presented EnDQT in \cite{Pipa2023ATheory}, but GQT shows how the idea for EnDQT arose and how it connects with other QTs. I will also show how GQT applies
to hybrid classical-quantum theories.\footnote{E.g., \cite{Oppenheim2023, Diosi1995QuantumLimit}.} I will then compare GQT with wave function realism and
the primitive ontology framework and argue that it provides important
benefits that these views don't provide, and without some of their costs
(section 2.4). Also, I will argue that it allows us to make a new
comparison between QTs and argue that EnDQT should be
preferred in a certain sense. 




According to GQT, systems can occupy spatiotemporal regions (ST version)
or give rise to spacetime (non-ST version) using, for example, an
appropriate theory of quantum gravity. I will focus on the ST version
for simplicity, but spacetime regions aren't necessarily fundamental in
this view. To simplify, I will mainly assume non-relativistic QT and the
Schrödinger picture Hilbert space-based finite-dimensional QT in the
presentation of the theories. Furthermore, note that given the
non-reificatory approach to quantum states, viewing it as only as an
auxiliary tool, GQT doesn't rely on a particular mathematical
formulation of QT, and it can be expressed in terms of other
formulations.\footnote{Such as the Heisenberg picture, interaction
  picture, and the Lagrangian formulation.}

\section{Some generative quantum theories based on quantum
properties}\label{some-generative-quantum-theories-based-on-quantum-properties}

I will start presenting a generative version of GRW, which I will call
generative-GRW. This theory will serve to explain in the following order
the basic features assumed by GQT: generators, generative properties,
the kinds of determinate values that generators generate, the ontology
of properties adopted, the conditions that establish how generators
account for determinacy, and two structural features that help explain
how determinacy arises via interactions.

\subsection{A generative collapse theory and introduction to
GQT}\label{a-generative-collapse-theory-and-introduction-to-gqt}
We can characterize the role of any ``interpretation'' of QT or QT as
giving an account of how systems end up having determinate values,
although, given the EEL, unitary interactions leave such values
indeterminate.

To give a general account of how systems come to have determinate
values, GQT introduces generator systems of determinacy or
\emph{generators}. Note that the word ``generate'' will be used in two
different senses. It will be used to designate how GQT helps build new QTs, and how some elements of GQT give rise to
determinacy. Generators are systems that have the capacity to give rise
to other systems having determinate values. \emph{Non-generator systems}
don't have the capacity to give rise to other systems having determinate
values. Also, we have \emph{generative properties}, which are the
properties that generators have via which they influence other systems
to have determinate values. A key claim of GQT is that each QT
introduces different generators and generative properties, which
generate different kinds of determinate values. Generators and
generative properties are two interrelated features that help generate
QTs. As it will become clearer, in the case of
generative-GRW, the generators are the systems that have positions, and
some of their positions are generative properties.

The kind of determinate values that each generator generates is another feature
that helps generate QTs. The determinate values generated
by generators can be absolute (i.e., don't vary according to systems),
relative to a system (for system $X$, system $Y$ has a determinate value v,
but for system $Z$, $Y$ has a value v' or an indeterminate value), relative
to multiple copies of systems (multiple copies of the same interacting
systems arise, each with different possible determinate values like in
the MWI), etc. Unless stated otherwise, I will consider the determinate
values generated as absolute, although in the next section, we will see
other possibilities.
GQT, in principle, allows for multiple ontologies of properties, and
this is another one of the features that help generate QTs (see
section 4). In this article, I will propose and focus on an ontology of
properties where it is manifest when systems $S'$, interacting with other
system S, have the properties that give rise to $S$ having a determinate
value, i.e., it becomes manifest when systems $S'$ have generative
properties.

According to this ontology, for GRW and GQT in general, systems are
collections of quantum properties, and value properties (henceforward,
values) are related to quantum properties (more on this below). To
explain what quantum properties are, first note that I will assume that GRW
considers that there are fundamental quantum systems called particles. I will consider that
particles are systems that have different subsystems, each a collection of certain
quantum properties. Only one of the systems has position quantum
properties (alongside momentum and energy), and the other systems have
other quantum properties, such as spin and energy. The former system, as
I have mentioned above, is a generator because it has
the capacity to give rise to other systems having determinate values
(having generative quantum properties). This contrasts with the
subsystem of the particle that has spin quantum properties, which isn't
a generator. Quantum properties of subsystems of particles are
represented via quantum states belonging to different Hilbert spaces and
self-adjoint operators (which I will call observables) that act on those
spaces.

Quantum properties have a feature called \emph{differentiation}, which
impacts the determinacy of the values that systems having those
properties give rise to. Interactions with generators change the degree
of differentiation of a quantum property that a target system and the
generators have (we will see further below why differentiation comes in
degrees). More concretely, the differentiation of a quantum property
that a target system $S$ might end up with due to an interaction with
generators can be inferred and measured via the distinguishability of
certain quantum states of the generators concerning the quantum states
of $S$ (hence the use of the term differentiation). Furthermore, the
quantum states of $S$ are eigenstates of an observable that also
represents that property.\footnote{Or improper eigenstates in the
  non-idealized case of systems whose quantum properties are represented
  via infinite-dimensional Hilbert spaces, such as position quantum
  properties. See, e.g., \cite{Wallace2019WhatMechanics}.} When such differentiation is
\emph{maximal} and \emph{stable} (in a certain sense to be defined
soon), I consider that we end up having $S$ with that quantum property
\emph{stably differentiated}, and $S$ will have a determinate value
related to that quantum property. Crucially, generators under
interactions with $S$ will also have a quantum property stably
differentiated, and thus a determinate value, when they give to $S$ having
a stably differentiated quantum property (again, to some degree at
least, as we shall see). This quantum property of the generators is a
\emph{generative quantum property}. So, for GQT,\\

\noindent All generative quantum properties of a generator system are fully
stably differentiated.\\

Note that not all stably differentiated quantum properties are
generative. Spin quantum properties in different directions in GRW are
never generative. It is the subsystem of the particle that has certain
positions that gives rise to other systems having a determinate value,
not the subsystem that has spin.

As we will see, the use of the term \emph{stable} is because the process
that allows us to infer if there is determinacy will often involve, via
decoherence, the analysis of a certain quantity that should assume a
stable value over time to license those inferences, i.e., the
distinguishability/differentiation of the quantum states of an
``environmental system'' interacting with a target system. So, I will
consider that the stabilization of the differentiation of a quantum
property of a target system $S$ in most QTs arises via a
\emph{stable quasi-irreversible} or \emph{irreversible process} that
gives rise to and fixes the determinacy of the value in some degree
proportional to the degree of differentiation of the quantum property
that $S$ has at some time t. More concretely, this process leads $S$ to have
a quantum property D*-P with a degree of differentiation D* to give rise
to a value of P (e.g., spin in different directions, position, etc.)
with a degree of determinacy D=D* at t. I will call the process in which
a generator system $S'$, having a generative quantum property, gives rise
to other systems $S$ having a stably differentiated quantum property, a
\emph{process of stable differentiation} or the stable differentiation
of quantum properties of $S$ by $S'$.

In generative-GRW, this process is the process of collapse or
spontaneous localization. It leads systems to have a stably
differentiated quantum property and, thus, a determinate value. In the
MWI and EnDQT this process is a quasi-irreversible and irreversible
process, respectively, represented via decoherence. In the former case,
it is called the process of branching into different worlds.

Let's turn to further details of this view. I will consider that\\

\noindent Unless stated otherwise, in the absence of interactions or other
processes that lead to a process of stable differentiation, quantum
properties of systems will be undifferentiated, which means having the
lowest degree of differentiation. So, some interactions or processes
change the differentiation of the quantum properties that systems have.\\

Thus, indeterminate values/undifferentiated quantum properties are the
default features of systems. Only under certain processes and
interactions do systems having determinate values arise. It is in this
sense that for GQT, quantum indeterminacy plays an important explanatory
role, being an important tool to interpret QTs: it
establishes when determinacy doesn't arise, being a default feature of
systems. The exceptions in this article are Bohmian mechanics, where
some systems have always determinate values of position, and hybrid
classical-quantum theories, which will have, for example, the metric and
its conjugate momentum always stably differentiated. In the case of
these two theories, having determinate values or indeterminate values
are both default features of systems.
Note that, perhaps, one may consider the (typically regarded)
non-dynamical quantum properties (e.g., electric charge, mass) to be
also stably differentiated by default. However, the QTs investigated
here can assume that ``non-dynamical'' observables represent
undifferentiated quantum properties that become stably differentiated
under interactions with the typical appropriate environments. Although
this is not mandatory when adopting GQT, for simplicity, I will assume
this in this article.\footnote{Decoherence was proposed to account for such so-called superselection rules (see, e.g., \cite{Earman2008SuperselectionPhilosophers, GIULINI1995291}) So decoherence by an appropriate environment
  could be used by at least by some QTs to represent and infer such
  interaction.}

\emph{Stability conditions} are the conditions under which a system
comes to have a stably differentiated quantum property or (more
generally) a determinate value, and they are another feature that helps
generate QTs. Stability conditions for generative and
non-generative quantum properties may differ in the case of the theories
explained in this section, and to summarize, they are the following,\\

\noindent A system $S$ has a stably differentiated quantum property, giving rise to
S having a determinate value associated with that property when,

\begin{enumerate}
\def\labelenumi{\roman{enumi})}
\item
  if that quantum property is a generative one (which only generators
  have), $S$ has that stably differentiated quantum property due to a
  spontaneous chancy collapse process (GRW), due to or in certain
  interactions (MWI, single-world relationalist views, and EnDQT), or S
  has by default that quantum property stably differentiated (Bohmian
  mechanics and hybrid classical-quantum theories); or
\item
  if that quantum property isn't a generative one, $S$ is interacting with
  generators that have a generative quantum property, which gives rise
  to $S$ having that stably differentiated quantum property.
\end{enumerate}

Regarding i), for example, Bohmian mechanics considers that the position
quantum properties of systems are always stably differentiated by
default, and those are the generative quantum properties. GRW considers
that generators can be subject to collapse, which gives rise to systems
having stably differentiated position quantum properties independently
of the interactions they have with other systems, going from having an
undifferentiated position to a stably differentiated one. More
concretely, systems in GRW often evolve unitarily; however, they have
the probability per unit time $\lambda$ of indeterministically being
localized, \emph{collapsing,} and having at least a stably
differentiated and determinate value of position. Note that I don't mean
that collapse refers to the wave function but the stable differentiation
process involving systems since the wave function here is not considered
a real entity.

The collapse of a system \(S_{i}\) in a spacetime region is represented
via the multiplication of the total wave function written in the position
basis by a narrow Gaussian wave packet in the position basis whose width
is \(\sigma\), which represents the localization accuracy. Moreover, the
probability of the wavepacket being centered in region $C$ is given by the
Born rule. The stably differentiated quantum property of the generator
system \(S_{i}\) affected by collapse is represented by the
post-collapse wave function\footnote{Given that the quantum state has an
  inferential role, I will accept the standard assumption that we can
  ignore the global phases of the quantum state to make inferences about
  and represent properties.} plus the observable position that acts on
the Hilbert space of system \(S_{i}\). The possible determinate values
of \(S_{i}\) are represented by the eigenvalues of the observable that
the position quantum states of \(S_{i}\) are eigenstates of.\footnote{Due
  to the continuous spectrum of the position observable, it brings some
  extra complications. However, given our finite-dimensional Hilbert
  space idealization, I will neglect them. See, e.g., \cite{Wallace2019WhatMechanics} for
  ways of dealing with it. Also, the wave function leaves some ``tails''
  upon collapse, assuming the representational and inferential role of
  the quantum states assumed by GQT, the approximate ways of
  representing determinate values aren't a problem for this view (more
  on this below).}

Regarding ii), generative-GRW also considers that when a generator or
non-generator target system $S$ interacts with a certain generator system
or systems $S'$, so that they get \emph{entangled}\footnote{I will make
  precise below the interactions represented via entanglement.} and a
collapse happens, this leads to the stable differentiation of the
quantum properties of $S$ by $S'$. More precisely, we can infer that there
is an interaction that involves generators $S'$ having a generative
quantum property, which leads the target system $S$ to have a determinate
value. The stably differentiated quantum property of the target system
will be represented by the eigenstates of the observable concerning a
property (spin in different directions, etc.) that are correlated with
the position states of the generator or generators upon collapse plus
that observable.
As I have said above, the full distinguishability of the quantum states
in a superposition of the generator or a collection of generators
constituting system $S'$ concerning the quantum states of the target
system $S$ (which could be a generator or not) \emph{just before} collapse
(or another process of stable differentiation in the case of other QTs)
allows us to infer which stably differentiated quantum property $S$ will
have due to $S'$ after a certain time. These quantum properties often go
beyond position and can be energy, spin in a direction, etc.

In the case of generative-GRW, the stable differentiation of a quantum
property of a target system (i.e., a system under analysis by a model)
can be inferred via the quasi-irreversible process of decoherence of the
target system by its environment composed of many generators, and which
occurs just before collapse. This is because quasi-irreversible
decoherence will typically require many environmental systems having a
position (which is correlated with the position of the others) to be
stably entangled over time with the target system. So, very likely, some
will collapse, triggering a collapse process that leads the others
within that environment to have a determinate value due to their
correlations, as well as the target system. More on this process below.

Relatedly, it is plausible to consider that at least some quantum
properties can be stably differentiated in terms of different degrees,
and this impacts the subsequent \emph{degree of determinacy} that arises
from those quantum properties. For example, in the double-slit
experiment, if the detectors at the slits interact with a quantum system
weakly, in such a way that we can't fully distinguish in which slit it
passed, we get some disappearance of interference. These interactions
will give rise to a low entanglement between the position and the
degrees of freedom of the detector. Furthermore, the more the
interactions between the target system and the detector distinguish the
path of the system, the more entanglement we have between the position
of the target system and the degrees of freedom of the detector, and the
more the interference tends to disappear until it disappears completely
under maximal entanglement. So, I will consider that quantum properties
come in terms of different \emph{degrees of (stable) differentiation},
as well as the determinacy of the resultant values.

For example, a system can have different quantum properties spin-x with
different degrees of differentiation over time. Values come in terms of
degrees of determinacy D and depend on the degree of differentiation D*
of quantum properties. A quantum property is undifferentiated when it
has the lowest degree of differentiation and differentiated when it has
the highest one. A value with the maximum degree of determinacy is a
determinate value, and with a minimum degree of determinacy is an
indeterminate value.\footnote{Remember that this assumption is not
  mandatory when adopting since GQT allows for different property
  ontologies.}

I will now show more concretely how we can infer the degree of
differentiation of the quantum property that a system, after
interactions, ends up with via the degree of entanglement of its quantum
states with its environment and decoherence.

The degree of differentiation of a quantum property of a system can be
measured via the non-diagonal terms of the reduced density matrix of the
system subject to decoherence when we trace out the degrees of freedom
of the environmental systems that are interacting or interacted with the
system of interest. Let's consider a toy scenario with system E, which
is a generator, constituted by many subsystems that interacted or are
interacting with system S. For instance, suppose $S$ has quantum
properties spin in different directions that are interacting strongly
(i.e., the Hamiltonian of interaction dominates the system's evolution
in the timescales of interest) with many systems with positions, which
constitute E.\footnote{Alternatively, in other QTs that
  don't privilege position, we could consider instead an environment
  with systems with spin in multiple directions. See, e.g., \cite{Cucchietti2005DecoherenceEnvironments}.}\footnote{Realist decoherence models involving
  environments with position quantum properties include, for example,
  collisional models of decoherence and models of quantum Brownian
  motion. See, e.g., \cite{Joos1985TheEnvironment, Kiefer1999Decoherence:Beyond, Schlosshauer2007DecoherenceTransition} and references therein.} For simplicity,
throughout this article, I will assume this kind of evolution of the
system under the interactions that lead to decoherence.\footnote{More complex models of decoherence (see, e.g., \cite{Zurek2003DecoherenceClassical} where the
  systems don't interact strongly with the environment, which involves
  the self-Hamiltonian having more weight on their evolution, may give
  rise to different observables with determinate values depending on the
  initial quantum states. More on this below.} Let's assume some
situations where $S$ initially has an undifferentiated spin-z quantum
property. $S$ then interacts with E, and their interaction is represented
via the standard von Neumann interaction at least approximately as
\({| \uparrow}_{z}\rangle_{S}|E_{0}\left( t \right)\rangle_{\text{E\ DS}} \rightarrow {| \uparrow}_{z}\rangle_{S}|E_{\uparrow}\left( t \right)\rangle_{\text{E\ DS}}\),
\({| \downarrow}_{z}\rangle_{S}|E_{0}\left( t \right)\rangle_{\text{E\ DS}} \rightarrow {| \uparrow}_{z}\rangle_{S}|E_{\downarrow}\left( t \right)\rangle_{\text{E\ DS}}\),
or as
\begin{equation}
    (| \uparrow_{z}\rangle_{S} + | \downarrow_{z}\rangle_{S})|E_{0}(t)\rangle_{\text{E DS}} \rightarrow \alpha| \uparrow_{z}\rangle_{S}|E_{\uparrow}(t)\rangle_{\text{E DS}} + \beta| \downarrow_{z}\rangle_{S}|E_{\downarrow}(t)\rangle_{\text{E DS}}
    \label{decoherence}
\end{equation}

The change in the degree of differentiation of the quantum property
spin-z of $S$ upon this interaction can be inferred and calculated through
the reduced density operator \({\hat{\rho}}_{S}\left( t \right)\), which
is obtained by doing the partial trace of the degrees of freedom of the
environment. More concretely, this analysis is done through the overlap
terms that concern the distinguishability of the states of $E$ with
respect to the spin-z of S, i.e.,
\({\langle E}_{\uparrow}(t)|E_{\downarrow}\left( t \right)\rangle_{\text{E\ DS}}\)
and
\({\langle E}_{\downarrow}(t)|E_{\uparrow}\left( t \right)\rangle_{\text{E\ DS}}\).
More generally, consider a system $S$ that initially has an initially
undifferentiated quantum property D*-P, where the observable that
concerns P has eigenstates \(\left| s_{i} \right\rangle_{S}\). Given the
interaction between $S$ and environmental system E, after tracing out the
degrees of freedom of E, we obtain that

\begin{equation}
\begin{aligned}
    \hat{\rho}_{S}(t) = \sum_{i = 1}^{N} |\alpha_{i}|^{2} |s_{i}\rangle_{S} \langle s_{i}| 
    + \sum_{\substack{i,l = 1 \\ i \neq l}}^{N} \Big( \alpha_{i}^{*}\alpha_{l} |s_{i}\rangle_{S} \langle s_{l}| \langle E_{i}(t) | E_{l}(t) \rangle_{\text{E DS}} \\
    + \alpha_{l}^{*}\alpha_{i} |s_{l}\rangle_{S} \langle s_{i}| \langle E_{l}(t) | E_{i}(t) \rangle_{\text{E DS}} \Big)
\end{aligned}
\end{equation}

Then, a measure of the degree of differentiation of the quantum property
D*-P of $S$ in the spatiotemporal region ST for the simple scenarios that
we are considering will be given by the von Neumann entropy\footnote{Given
  a density operator \(\rho_{S}\ \)for quantum system S, the von Neumann
  entropy is\(\ S(\rho_{S}) = - tr(\rho_{S}\ln\rho_{S})\).
  \(S({\hat{\rho}}_{\text{S\ }})\) is zero for pure states and equal to
  \(\text{ln\ N}\) for maximally mixed states in this finite-dimensional
  case.} \(S\left( {\hat{\rho}}_{S}(t \right))\) of
\({\hat{\rho}}_{S}\left( t \right)\) over \(\text{lnN}\), where \(N\) is
the number of eigenvalues of \({\hat{\rho}}_{S}\left( t \right),\)

\begin{equation}
D^{*}(P, S, ST, t) = \frac{S(\hat{\rho}_{S}(t))}{\ln N}
\end{equation}

If $D^*(P,S,ST,t)$ via the above overlap terms goes
quasi-irreversibly, \emph{i.e., stably,} to one over time (in the sense
that the recurrence of this term back to not being significantly
different from zero is astronomically large), and these interactions
involve many environmental systems that make this process hard to
reverse, it is considered that $S$ is decohered by E. In the QTs that
appeal crucially to decoherence to infer when systems have determinate
value, such as the MWI (Section 2.2) and EnDQT (Section 2.4), it is
inferred that when decoherence occurs, $S$ has a \emph{stably}
differentiated quantum property, having a determinate value due to E
(but with some caveats in the case of EnDQT). More precisely, we can
infer from this process that $E$ also has a stably differentiated quantum
property/generative quantum property, which leads $S$ to also have a
stably differentiated quantum property.

Upon knowing the actual result, we update the state of $S$ to one of the
\(\left| s_{i} \right\rangle_{S}\), and consider that the system has a
determinate value, which is an eigenvalue of the observable that
\(\left| s_{i} \right\rangle_{S}\) is an eigenstate of. Similarly, for
E, where its possible determinate values will be the eigenvalues of the
observable that \(|E_{i}\rangle_{E}\) are eigenstates of. In the
language typically employed by decoherence theorists,
\(\left| s_{i} \right\rangle_{S}\) for each \(i\) are pointer states,
and the observable that these states are eigenstates of is the pointer
observable ``selected'' by the environment E.\footnote{Note that pointer
  states here don't necessarily refer to the quantum states of a
  measurement device, but whatever is the target system.} In GRW, also
taking into account the account the collapse laws, we can infer there
will be a collapse to when this occurs with an environment constituted
by systems with the quantum property position. However, the collapse
timescale is typically longer than the decoherence timescale.\footnote{See
  \cite{sep-qm-decoherence} and references therein for the relation between
  collapse theories, decoherence, and their timescales.}

More generally, we can measure and represent the degree of
differentiation D* of the quantum property D*-P that $S$ will end up with
at the end of the interaction with $E$ at t, with $0\leq D^*(P,S,ST,t) \leq 1$,  in the possible
elements of the set of spacetime regions ST where $S$ is differentiated by
E. At least in the case of the MWI and EnDQT, we can also infer the
differentiation timescale, which is equal to the decoherence timescale.
This is done by analyzing the value in which
$D^*(P,S,ST,t)$ stably converges over time.

Thus, note that, as I have mentioned, a quantum property of $S$ might not
be fully stably differentiated and just be \emph{stably differentiated}
to some degree D* by E, and thus, it gives rise to a value with a degree
of determinacy D = D*. This happens if the above quantum states of the
environment have a certain \emph{stable} non-zero overlap over time
(notice how stability plays a role in these inferences). So, it is
considered that in order for generators to have a generative quantum
property and hence give rise to this process, they need to give rise to
a quasi-irreversible process, which involves many degrees of freedom of
the environment, in such a way that they decohere the target system
\emph{to some degree}.

The decoherence in these scenarios gives rise to the following
criterion: in order for system $S$ to have a determinate value v of
\(O_{S}\), the observable \(O_{S}\) of $S$ that is monitored by system E,
and whose eigenstates are decohered by $E$ in the sense above, has to at
least approximately commute with the Hamiltonian of interaction
\(H_{\text{SE}}\) representing the interaction between $S$ and E, i.e.,
\(\left\lbrack H_{\text{SE}},O_{S} \right\rbrack \approx 0\). This is
the so-called commutativity criterion.\footnote{See \cite{Schlosshauer2007DecoherenceTransition} and references therein. This criterion implies that all terms in a
  Hamiltonian of interaction will individually satisfy this criterion.
  In more complex models of decoherence where the Hamiltonian of
  interaction doesn't dominate the evolution of the systems, note that
  this monitoring may be indirect, such as the decoherence of momentum
  in more complex models of decoherence than the ones mentioned here
  \cite{Zurek1993CoherentDecoherence}, where there is direct monitoring of the
  position. The latter is contained in the Hamiltonian of interaction of
  the system (but not the former), and that's why the decoherence of the
  momentum is indirect.} The value $v$ is among the possible eigenvalues of
\(O_{S}\).

We can use decoherence to represent quantum properties. The generative
(stably differentiated) quantum property of the target system is
represented by the quantum states in the superposition that are
decohered by (or entangled with) the generator plus the observable that
these quantum states are eigenstates of. The generative (stably
differentiated) quantum property of the generator is represented by the
quantum states that decohere the quantum states of the target system to
some degree and the observables that such quantum states are eigenstates
of.

However, not all interactions with generators\footnote{Or, at least in
  the case of EnDQT and MWI, with systems that could end up being
  generators.} give rise to systems having a determinate value, although
there is something that changes in the quantum properties of the systems
under these interactions. Consider the spin of a particle in different
directions in a series of Stern-Gerlach devices without letting the
particles hit a screen between each device. This leads the system
\(S^{*}\) with a spin in a certain direction to interact with the
generator \(S^{'}\), leading to their entanglement. Assuming the GRW
theory, there is something that changes in the spin direction of the
quantum systems when they go from one magnet to the other, but (very
likely) there is no collapse/stable differentiation. If there were, we
would have an irreversible process, and thus, we wouldn't be able to
reverse the result of the operations by having a Stern-Gerlach
interferometer that reverses the state of the particle to its previous
state. So, it is plausible to consider that the spin of the system that
interacts with the generator has an indeterminate value, although there
is something that changes in the quantum property that corresponds to
that indeterminate value.

In most QTs presented here, the interactions that don't
lead to stable differentiation, such as the one above, can be
\emph{inferred} and represented simply by the quantum states and
observables in the models where we have entanglement between the degrees
of freedom of interacting quantum systems,\footnote{QTs will often
  postulate different structures that establish when systems are
  interacting or not. I will come to that soon.} or relatedly where we
have the so-called virtual/\emph{reversible} decoherence. This
decoherence involves ``entangling'' interactions that are reversible,
not giving rise to an irreversible or quasi-irreversible physical
process, because often they don't involve enough environmental systems
that make such process hard to reverse unitarily, and thus, it is not
typically considered real/irreversible decoherence. In the case of GRW,
this reversible process involves the entanglement between the quantum
states of a small number of generator or generators
\({S_{\ }^{'}}_{\ }\)in the position basis (the environment) and the
generator or non-generator target system \(S^{*}\).\footnote{See, e.g., \cite{Oliveira2006CoherenceAE}.} Taking into account the collapse
laws, since it doesn't involve sufficient systems to very likely
collapse occurs, it allows us to infer that stable differentiation likely
won't occur. In the Stern-Gerlach case above, we obtain that both
systems, after interacting, are represented by

\begin{equation}
|\Psi(t')\rangle = \frac{1}{\sqrt{2}} \left( |\uparrow_{z}\rangle_{S^{*}} |\text{up}\rangle_{S'} + |\downarrow_{z}\rangle_{S^{*}} |\text{down}\rangle_{S'} \right)
\end{equation}
and the self-adjoint operators spin-z and position that act on the
position and spin Hilbert spaces of S. So, via entanglement and
reversible decoherence, we represent and infer those systems that have
an undifferentiated quantum property in interactions with other systems,
and thus, when generators don't have generative quantum
properties.\footnote{Note that, if we had collapse, the quantum property
  of the non-generator system would be represented either by
  \({| \uparrow}_{z} >_{S^{*}}\) or \({| \downarrow}_{z} >_{S^{*}}\)
  plus the spin-z observable. The determinate values that arise from the
  spin-z quantum property are represented by \(\uparrow_{z}\) or
  \(\downarrow_{z}\).}
So, GQT considers decoherence as an epistemic tool that can
be used to infer which systems have undifferentiated or stably
differentiated quantum properties and to infer if they will give rise or
not to stable differentiation and, therefore, determinacy through
interactions. Equivalently, it serves as a tool to infer if generators
will have a generative quantum property under interactions. Furthermore,
for some QTs (such as in the MWI) decoherence also allows us to infer
which systems are generators and have generative quantum properties.
Some factors will need to be taken into account in the use of
decoherence as an inferential tool, some of them already mentioned
above, but I want to emphasize them.
To infer if we have systems with undifferentiated or stably
differentiated quantum properties, we need to find if we have a reversible or an
irreversible process of decoherence, respectively. This analysis largely appeals to
pragmatic factors, such as the number of systems that interact with the
target system. It is also necessary to analyze the quantum properties of
the systems of the whole environment that interact with the target
system during different times since specific environmental systems may
contribute more to determining the degree of differentiation of the
quantum property that the system under analysis might end up with. For
example, at least in the case of GRW, in the Stern-Gerlach apparatus,
the (reversible decohering) interaction between the spin and the
position degrees of freedom of the particle crucially contributes to
measuring the degree of differentiation of spin-z, \emph{but not} the
stability of that quantum property and the determinacy that it arises.
Afterward, the degrees of freedom of the particles that constitute the
screen detector can be regarded (let's assume that there is a collapse at
the screen) as contributing to the particle having a stably
differentiated position, which ends up leading it to also have a stably
differentiated spin. So, the degree of the stably differentiated spin
that the particle ends up with depends on the interaction that started
previously with the subsystem of the particle that has the quantum
property position. Nevertheless, depending
on the context, note that both reversible and irreversible decoherence
allow us to measure the degree of differentiation of a quantum property
via the degree of entanglement/distinguishability of the quantum states
of the environment that are correlated with the quantum states
concerning that quantum property. I will make this idea more precise below by
distinguishing different kinds of interactions. 

Finally, for some QTs, it is necessary to analyze whether decoherence
involves generators that very likely will have generative quantum
properties, giving rise to the target system having a quantum property
stably differentiated to some degree. As we saw in the GRW case, this
should be an environment that collapses the target system in such a way
that \emph{it distinguishes} its different quantum states that were
previously in a superposition.

As mentioned before, differentiation and determinacy are related, and
this allows for an analysis of quantum indeterminacy and determinacy.
This relation will establish that a property P*, in this case, a value
property, is the property of having some other property P** having
specific features. So, I will consider that\\

\noindent \emph{For a system} \emph{to have a value v of P (where P could be
energy, position, etc.) with a non-minimal degree of determinacy D is to
have stably differentiated quantum property D*-P to a non-minimal degree
D* where D=D*. A system with a quantum property (fully) stably
differentiated will have a determinate value of P. }\\

On the other hand, indeterminacy and undifferentiation are related,\\

\noindent \emph{For a system to have an indeterminate value of P is to have an
undifferentiated quantum property.}\\

Note that according to this relation, we have \emph{multiple} quantum
properties concerning P, represented by quantum states and observables,
that correspond to a non-maximally determinate value of P.\footnote{Note
  also that this relation doesn't imply that undifferentiated quantum
  properties are more fundamental than indeterminate value properties.}
Just think about the variety of eigenstates of an observable concerning
P that we can superpose and/or entangle with the quantum states of other
systems to get quantum states that allow us to represent an
indeterminate value of P.

Now that I have presented the ontology of properties that GQT will adopt
in this article, I will turn to two structural features, which will help
to give rise to different QTs. Certain structures, which
include different kinds of interactions, account for how determinacy
arises or not. Importantly, \emph{what constitutes an interaction} how
to infer it, and the different interactions that belong to the structure
of interactions, varies according to the QT. Kinds of interactions
between systems with undifferentiated quantum properties form structural
features called Indetermination Structures (ISs), where these
interactions don't involve or give rise to any system having a
determinate value. ISs are one of the structural features assumed by
GQT.
In the case of GRW, what I will call \emph{collapse-ISs} are represented and
inferred via equations such as \ref{decoherence}. Systems that don't belong to ISs
belong to Determination Structures (DSs), which also involve
different kinds of interactions between systems. I will call them
\emph{structural generators} since they give rise to determinacy. I will
call the DSs for generative-GRW, collapse-DSs. DSs is the last feature
considered by GQT that I will present in this article. As will become
clearer, each QT may adopt different DSs, ISs, property ontologies,
generators, stability conditions, generative properties, and kinds of
determinate values that generators generate. Also, as it can be seen, these
seven features are related to each other.

DSs and ISs can have a structure that may sometimes be represented by
directed graphs, undirected graphs, or a hybrid (thus being
structures). Nodes represent systems, and edges between nodes represent
certain kinds of interactions. One of them is \emph{Stable
Differentiation Interactions} (\emph{SDIs}), which involves an arrow
that goes from the generator or generators to the target system, leading
them to have determinate values.
On the other hand, we have Unstable Differentiation Interactions (UDIs), which are a sub-kind of ISs. UDIs are interactions between
systems $S'$ and $S''$ in which \emph{if} some generator $S$ stably
differentiated a quantum property of $S'$/S'', it would also stably
differentiate a quantum property of $S''$/S' to a degree inferred from how
much the quantum states of $S'$/S'' distinguish the quantum states of
S''/S' (or in other words, how much the quantum states of $S'$/S'' are
distinguishable). For instance, in the Stern-Gerlach case above, if some
environment $E$ in the screen stably differentiates the position quantum
property of system $S''$ (which is a subsystem of the particle S), the
spin-z of $S'$ (another subsystem of S) is stably differentiated to a
degree D* that is quantified by the overlap of the quantum states in the
position basis of $S''$ that are entangled with the states of $S'$. Also, we
have UDIs where one or both systems are generators, and if a quantum
property of $S'$/S'' became stably differentiated, a quantum property of
S''/S' would also become stably differentiated, where such degree of
differentiation would be measured like in the above case. As I will
explain, UDIs might have a direction.

UDIs can be inferred via reversible decoherence with no collapse like
the one we have seen above, or simply when we have entangled systems.
So, as we can see, unstable differentiation interactions don't give rise
to an irreversible qua stable process, but instead to a \emph{reversible
qua unstable process}. Therefore, they don't change the stable
differentiation of quantum properties that systems have although they
could end up leading to processes that change it as I have explained
above.

I will now introduce other interactions between systems that belong to
ISs with an example that shows how generative-GRW accounts for
interference phenomena. I will also demonstrate some extra explanatory
resources that GQT allows for in accounting for interference phenomena,
although builders of generative quantum theories might not wish to
assume them.\footnote{They may wish to not introduce the interactions
  that I will introduce below and ``systems having different
  locations.'' However, this will likely diminish their explanatory
  resources.}

I will consider that systems can occupy multiple ``locations,'' allowing
us to represent the relations of influence behind interference
phenomena, but without appealing to the wave function. The trick is to
use the interactions that DSs and ISs allow for. When systems have an
indeterminate position value, they are associated with multiple
locations, and we can call each system-location pair a ``part'' of the
system, and these parts in these multiple locations interact via
\emph{potential destruction interactions}. So, the latter are
self-interactions that systems develop between the different parts of
themselves that occupy different regions of spacetime. These
interactions also belong to collapse-ISs, being reversible.

Relatedly, collapse-DSs also involve self-interactions called
(\emph{actual)} \emph{destruction interactions}, and they arise from the
potential ones. This interaction arises when one part of the system has
a quantum property stably differentiated, leading the system in the
other locations to not exist anymore (irreversibly). So, it occurs when
a system goes from having an indeterminate position to a determinate one
Note that when a potential destruction interaction turns into actual
destruction interaction, we have the phenomenon typically called
\emph{collapse of the wave function} in ontologies that reify it. Let's
see how this works by considering a system that goes through a
Stern-Gerlach interferometer with a detector placed in one of the arms.
Let's, for example, assume that we have a neutron $S$ constituted by
system \(S_{\ }^{'}\) having, among others, the quantum property
position and system S* having, among others, the quantum property
spin-x, which initially are stably differentiated when the electron is
prepared. When it reaches the first beam splitter, the system is split
into two locations, having an undifferentiated position and spin-z. So,
between the two locations, it is indeterminate where \(S_{\ }^{'}\) is.
Undirected potential destruction interactions are developed between the
parts of the system at these locations. They are undirected interactions
because they don't have any direction of influence. Also, \(S_{\ }^{'}\)
and \(S^{*}\) develop a directed UDI, since \(S_{\ }^{'}\) could end up
stably differentiating \(S^{*}\), but not vice-versa. The particle's
quantum state is the one of eq. \ref{decoherence}.

When the system interacts with a detector placed in one arm of the
interferometer, the energy of the particle is stably differentiated by
this detector, where the quantum state just before collapse is

\begin{equation}
\begin{aligned}
    |\Psi(t'')\rangle = \frac{1}{\sqrt{2}} \big( 
    &|\uparrow_{z}\rangle_{S^{*}} |\textit{up}\rangle_{S'} |E_{\text{detected D1}}\rangle \\
    &+ |\downarrow_{z}\rangle_{S^{*}} |\textit{down}\rangle_{S'} |E_{\text{Not detected D1}}\rangle 
    \big).
\end{aligned}
\label{detection}
\end{equation}

So, just before the collapse, the detector has for very brief moments an
indeterminate location of its pointer. Let's suppose that system $S$ is
stably differentiated by D1. $S'$ in the other branch of the
interferometer will ``destroyed.'' We would obtain
\(|\Psi' > \approx |\)
\(\uparrow_{z} >_{S^{*}}|\text{up} >_{{S_{\ }^{'}}_{\ }}{|E}_{{detected\ D1}_{\ }} >_{\ }\),
with $S$ having determinate value \(\uparrow_{z}\) and \(\text{up}\), and
the rest of the systems that constitute the detector having determinate
values correlated with these ones.

Note that the structure of the destruction relations is not directly
represented via the quantum state, but rather inferred from it and
represented via the directed graphs (more on this below). Note also that
although the state of the whole system after the collapse is not an
eigenstate of position, this is unproblematic because of the
non-literalistic representational role quantum states have for GQT. The
system being close to being in a quantum state associated with these
properties is enough to represent them.

Let's now consider an EPR-Bell scenario,\footnote{\cite{Bell1964OnParadox}.} where
space-like separated Alice and Bob perform random measurement on systems
in a singlet-state, giving rise to correlations. To account for
EPR-Bell-like correlations, we can also use DSs and ISs. Consider the
state below, representing particles \(S_{A}\) and \(S_{B}\) before
either Alice or Bob measuring them,

\begin{equation}
\begin{aligned}
    |\Psi(t)\rangle = \frac{1}{\sqrt{2}} \big( &|\uparrow_{z}\rangle_{A} |\downarrow_{z}\rangle_{B} + |\downarrow_{z}\rangle_{A} |\uparrow_{z}\rangle_{B}\big) \\
    &|R\rangle_{E_{A}} |R'\rangle_{E_{B}}  |R''\rangle_{L_{A}} |R''\rangle_{L_{B}}
\end{aligned}
\end{equation}

Above we have two systems, \(E_{A}\) and \(E_{B},\) with position R and
R', respectively, and two systems $A$ and B, each with an undifferentiated
spin in all directions. \(L_{A}\) and \(L_{B}\) are the measurement
devices of Alice and Bob before interacting with their target systems.
Taking into account the above entangled state, it is considered that the
structure of the IS is composed by systems \(A\) and \(B\) connected by
an undirected (non-local) UDI. It is undirected because it can go both
ways when one of the systems' spin in a direction becomes stably
differentiated.

\begin{figure}[H] 
    \centering
    \includegraphics[width=3.69608in,height=3.67741in]{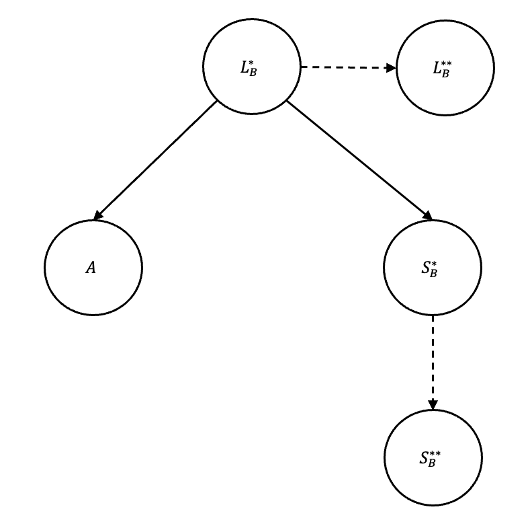}
    \caption{Directed graph representing a collapse-DS. Dashed arrows are
    destruction relations with a possible direction. The system arising in a
    determinate location (I have represented the part of the system at this
    location with a star) leads to the disappearance of the system being in
    the other location represented via a double star. The rest of the arrows
    are stable differentiation interactions.}
    \label{fig:collapseDS} 
\end{figure}

Afterward, it can happen (for example) that, in a certain reference
frame, \(L_{B}\) and \(E_{B}\ \)interact first with B, and we obtain the
following quantum state just before collapse,

\begin{equation}
\begin{aligned}
    |\Psi(t)\rangle = \frac{1}{\sqrt{2}} \big( &|\uparrow_{z}\rangle_{A} |\downarrow_{z}\rangle_{B} |{down'}\rangle_{E_{B}} |{down''}\rangle_{L_{B}} \\
    &+ |\downarrow_{z}\rangle_{A} |\uparrow_{z}\rangle_{B} |{up'}\rangle_{E_{B}} |{up''}\rangle_{L_{B}}\big) \\
    &|R\rangle_{E_{A}} |{R'''}\rangle_{L_{A}}.
\end{aligned}
\end{equation}

\(L_{B}\) will very likely have a stably differentiated position and
trigger a collapse process, which stably differentiates the quantum
properties of \(E_{B},\ B,\) and \(A\), and leads the potential
destruction relations that arose to become destruction relations. Below
(Figure \ref{fig:collapseDS}), we can see a directed graph representing the structure of
the DS that is formed.

To summarize, in this section, we have seen seven interrelated features
postulated by GQT: a property ontology, generators, generative
properties, the kinds of determinate values that generators generate,
stability conditions, DSs, and ISs. As it will become clear, the
combination of different sets of these features helps generate different
QTs.

\subsection{The generative-MWI and generative-single-world-relationalism}
Let's turn to different versions of the generative-Many-Worlds
Interpretation (MWI), and single-world relationalist views. Unlike
generative-GRW and, as we will see, generative-Bohmian mechanics, these
views don't necessarily consider that particles play a fundamental role.
Like in generative-GRW, all systems have indeterminate values by
default. Now, certain interactions give rise deterministically to
multiple copies of systems, each with the different possible stably
differentiated quantum properties and determinate values, giving rise to
what is typically called different worlds corresponding to different
sets of determinate values. Each world is represented by one of the terms in a
given superposition. Unlike generative-GRW, in principle, systems with
diverse quantum properties are generators, not just systems with
positions. So, many kinds of systems have the capacity to give rise to
determinate values individually or at least collectively.

In this case, the process of branching into different worlds is the
process of stable differentiation. The pattern explained in the previous
section is repeated here: when generators have generative quantum
properties, they can stably differentiate the quantum properties of the
other systems and thus lead them to have determinate values. However,
generative-MWI adopts stability conditions (Section 2.1.) that postulate that, in order
for the target system to have a stably differentiated quantum property,
it has to suffer a quasi-irreversible process due to its interactions
with generators having a generative quantum property. This process is
represented and inferred via the irreversible process of decoherence,
where generators decohere the target system. Relatedly, like I have
explained in the previous sections, decoherence is used to represent and
infer which systems are generators, and the properties of generators
that are generative and hence stably differentiated, giving rise to this
process. On the other hand, as I also have explained in the previous
section, via entanglement and reversible decoherence, we represent and
infer those systems that have an undifferentiated quantum property in
interactions with other systems, and thus, when systems don't have
generative quantum properties.

I will present the different generative-MWI views via examples in which
we have a Bell scenario where Alice (Lab A) and Bob (Lab B) can measure
their systems in only two possible directions. I will start with a
version of MWI where there is ``local'' branching,\footnote{See \cite{Sebens2018Self-locatingMechanics} for the distinction between local and global
  branching.} calling it \emph{generative-quasi-local-MWI} for reasons that I will soon explain. For
heuristic reasons, I will put a subscript DS in the quantum states of
systems that are generators and will have a generative quantum property
in the interactions under analysis, and thus give rise to interactions
belonging to a DS, stably differentiating other systems' quantum
properties. Furthermore, systems with different subscripts will belong
to different \emph{quasi-local-MWI-DSs}.

So, consider the following state,

\begin{equation}
\begin{aligned}
    |\Psi(t)\rangle_{A+B} = \frac{1}{\sqrt{2}} \big( &|\uparrow_{z}\rangle_{A} |\downarrow_{z}\rangle_{B} - |\downarrow_{z}\rangle_{A} |\uparrow_{z}\rangle_{B} \big) \\
    &\quad |E_{\text{ready}}\rangle_{\text{Lab A DS}'} |E'_{\text{ready}}\rangle_{\text{Lab B DS}''}
\end{aligned}
\end{equation}

Like in generative-GRW, we have UDIs involving $A$ and B. When Bob
interacts with his system, he stably differentiates the spin-z quantum
property of system B, which also leads to the non-local stable differentiation of
the spin-z of A. Such determinate values result in two worlds or more
precisely, two new quasi-local-MWI-DSs,
\begin{equation}
\begin{aligned}
    |\Psi(t')\rangle = \frac{1}{\sqrt{2}} \big( &|\uparrow_{z}\rangle_{A} |\downarrow_{z}\rangle_{B} |E'_{\downarrow_{z}}\rangle_{\text{Lab B DS}'} \\
    &- |\downarrow_{z}\rangle_{A} |\uparrow_{z}\rangle_{B} |E'_{\uparrow_{z}}\rangle_{\text{Lab B DS}'} \big) \\
    &|E_{\text{ready}}\rangle_{\text{Lab A DS}''}
\end{aligned}
\end{equation}
Note that Bob doesn't affect the branching of Alice. Only when Alice
interacts with $A$ will she branch into two other worlds, obtaining two
determinate values. Those values are only shared between the different
versions of Alice and Bob if they meet. Since there is some non-locality in this version of the MWI, I have called it quasi-local.

Note also that, before the above interaction, Lab B can have a stably
differentiated quantum property (being represented by
\(|{E^{'}}_{\text{ready}}\rangle_{\text{Lab\ B\ D}S^{''}}\) plus the
observable that this state is an eigenstate of) due to interactions that
Lab B is developing with other systems not included in the model. So,
systems that start having stably differentiated quantum properties may
persist in having stably differentiated quantum properties through
interactions, forming different worlds. The same in the case of Lab A.

As I have mentioned, GQT allows us to play around with the kinds of
determinate values that are generated. Instead of giving rise to
multiple determinate values deterministically, we could have a theory
like the above one, but generators give rise indeterministically to
single relative determinate values, which are relative to the different
generator systems,\footnote{See also the beginning of the previous
  section.} and $A$ would not be connected with $B$ via a UDI. So, the
stable differentiation of properties of $A$ would not affect the one of B,
and vice-versa. Furthermore, for Alice (if they don't interact), Bob and
his system would have indeterminate values, and vice-versa. This
\emph{generative-single-world-relationalism} resembles in some ways
other relationalist theories, such as Relational Quantum Mechanics and
Healey's pragmatism.\footnote{\cite{Rovelli1996RelationalMechanics, DiBiagio2021StableFacts, Healey2017ThePhilosophy}.}

Furthermore, briefly, if we wanted a
generative-single-world-relationalist theory that resembles more
Relational Quantum Mechanics (RQM), we would consider that any system is
a generator and gives rise to/generates relative determinate values upon
\emph{any} interaction. So, all quantum properties assumed by systems
under interactions are generative. However, when not interacting, it is
considered that systems have indeterminate values relative to each
other. Then, we would consider that the role of decoherence is to infer
when relative determinate values of a target system $S$ and records of
those values are inevitably shared between (environmental)
larger systems $S'$, $S''$, $S'''$, $S''''$, etc. that also interact, where $S'$ locally
interacts first with S, decohering it. Then $S''$ interacts with $S'$, gets
entangled with $S$ and $S'$, and obtains a record of the determinate value
of S, and so on for $S'''$ that interacts with $S''$, etc. In other words, decoherence is used to infer when systems inevitably stably differentiate the quantum properties
of each other, giving rise to shared relative determinate values
concerning the systems they interact with, or records of determinate
values. The above chains of local interactions would be the DSs for
\emph{generative-RQM}.

GQT also allows us to play around with the structure of DSs and ISs and
generate other generative-MWIs. For instance, in one generative-MWI we
could have an IS where when Bob interacts with his system, he leads $B$ to
have a stably differentiated spin-z quantum property, \emph{but this
doesn't} lead $A$ to have a stably differentiated spin-z, and vice-versa,
and so we wouldn't have the above UDIs. This renders the MWI local in
the sense that there is no influence between Alice and Bob in so far
that they can be considered space-like separated.\footnote{There are
  issues here regarding how we can determine space-like separation if
  Alice and Bob don't share the same world. I will discuss a related
  issue in Section 3} Let's call this version generative-local-MWI because it doesn't have the non-locality that I have identified above in the quasi-local version.
Another generative-MWI would consider that instead of preexistent
non-local UDIs, we have a theory with SDIs leading to non-local
interactions between systems having different stably differentiated
quantum properties over time. These SDIs would establish which systems
belong to the same world. Let's call it \emph{generative-global-MWI}.
Bob and Alice would be connected via an SDI, and the splitting into
branches of Bob when he measures his system would non-locally split
Alice into multiple worlds even before she does her measurement or
vice-versa. The resultant worlds of this branching would each contain
different systems connected by SDIs that establish if they belong to the
same world.

\subsection{The generative-Bohmian
mechanics}\label{the-generative-bohmian-mechanics}

Like generative-GRW, \emph{generative-Bohmian mechanics} considers
systems with quantum properties position as generators. The positions
and velocities will be generative quantum properties,\footnote{Why
  assume that these are the ones? We can point to, for example, to these
  being the only ones present in all interactions that give rise to
  determinate values, having an important explanatory role in the
  theory.} being, by default, stably differentiated. The rest of the
quantum properties will lead to a behavior similar to GRW, but without
irreversibility since the theory is deterministic. Like in
generative-GRW, we have fundamental particle quantum systems. The
guiding equation represents the velocity of the particles with a stably
differentiated position and how the latter changes over time, where this
equation depends on the quantum states of systems.

The degree of decoherence or entanglement between the quantum states of
the target system and the wave function of the particles in the position
basis allows for a measure of the degree of stable differentiation of a
quantum property of the target system upon interactions with these later
systems. For instance, in the case of spin in a certain direction, the
stable differentiation is measured via the overlap of the wave function
in the position basis, where such wave functions distinguish the
eigenstates of spin in that direction. The stable differentiation of
other quantum properties, such as the energy of the particle, can be
measured via the decoherence of the particle wave function by its
environment constituted by systems that have the position quantum
property.

To present generative-Bohmian mechanics in more detail, I will go over
examples. Bohmian mechanics, being a hidden variable theory, also leads
to the interpretation of quantum states as concerning our ignorance
about which quantum properties of the particle are stably
differentiated. Let's consider the one-particle case in a certain
Stern-Gerlach interferometer experiment. In the beginning, we have a
particle constituted by two subsystems, one subsystem $A$ with an
undifferentiated spin-z and a subsystem \(E_{A}\ \)with a stably
differentiated position within a larger region \(R_{\ }\),

\begin{equation}
|\Psi(t)\rangle = \frac{1}{\sqrt{2}} \left( |\uparrow_{z}\rangle + |\downarrow_{z}\rangle \right) |R\rangle_{E_{A}}
\end{equation}

The eigenstate of the position of \(E_{A},\) \(|R \rangle_{E_{A}},\) concerns
our ignorance about the current value of the position of a particle within that larger region $R$.
Like in collapse theories I will proceed in a
different way to account more satisfactorily for interference. This
version of generative-Bohmian mechanics considers that it is associated
to each particle a so-called \emph{partner particle}. Partner particles
are systems with the quantum property position and behave like those
systems in GRW, having different ``parts'' in different spatiotemporal
locations, but now we don't have irreversible destruction relations.

Partner particles will play the role of the branches of the wave function
(including the empty branches, i.e., the branches that don't have
particles) and account for interference without reifying the
wave function, although they are inferred via it. Instead of having a
particle ``carried by a wave,'' we rather have a particle interacting
with its partner particle. Like other quantum properties in
generative-Bohmian mechanics, the position of a partner particle can be
undifferentiated or stably differentiated to a degree where the degree
of differentiation is measured via the amount of irreversible
decoherence that the wave function associated with the particle and the
partner particle suffers caused by the interaction with an environment.
Also, when the wave function of the partner particle is in an eigenstate
of the position operator, it has a determinate value of position, which
will coincide with the one of its associated particle.

Let's then continue with our Stern-Gerlach interferometer example. Let's
consider a system that passes by a Stern-Gerlach device, giving rise to
a particle in the arms of the interferometer that has a stably
differentiated position and spin-z as subsystems, and a partner particle
with an undifferentiated position (no irreversible decoherence is
involved). We are ignorant about the determinate value of the spin-z of
the particle because we are ignorant about the initial
conditions/position of the particle that entered the interferometer.
Like in collapse theories, the two locations of the parts of the partner
particle are interacting via potential destruction interactions when its
position is indeterminate. We represent the state of this particle and
its partner particle via

\begin{equation}
|\Psi(t')\rangle = \frac{1}{\sqrt{2}} \left( |\uparrow_{z}\rangle_{B} |up\rangle_{E_{\text{B DS}}} + |\downarrow_{z}\rangle_{B} |down\rangle_{E_{\text{B DS}}} \right)
\end{equation}

This interaction turns into a destruction interaction when one of the
parts of the partner particle has a stably differentiated position due
to, for example, a larger system such as a measurement device. However,
contrary to collapse theories, the destruction interaction can be
reversed (after a long time) to
a potential one.

If the interferometer is set in the appropriate way, the particle and
its partner can give rise to interference.\footnote{Note that the stable
  differentiation of the spin quantum property here is more easily
  reversible.} The degree of differentiation of the rest of the quantum
properties (beyond position and spin) depends on how differentiated they
are by systems with the quantum property position. If in the above
situation, we measure the system by placing a detector at one of the
arms of the interferometer, the interaction between the system and the detector gives rise to (for
example)  the particle having a stably differentiated energy. Also, as
I have said it will stably differentiate the position of the partner
particle, and the other part of
the partner particle that also goes through the other side will
disappear. Thus, we can update the wave function of the systems to an
(effective) wave function of the system, which represents the particle
and its partner particle with determinate values.

Let's see what the SDIs and ISs are for two particles with a stably
differentiated position and undifferentiated spin in any direction. To
do that, let's consider the again the EPR-Bell scenario with the quantum
systems prepared at the source in the state of eq. \ref{detection}, but let's ignore
the measurement devices of Alice and Bob. We have two systems, \(E_{A}\)
and \(E_{B},\) with stably differentiated position, which together with
systems $A$ and B, each with an undifferentiated spin in any direction,
constitute two particles \(S_{A}\) and \(S_{B},\) respectively.

The non-local structure of ISs is at least represented and inferred via
the entangled states between systems and other equations of Bohmian
mechanics. Subsystems of the particle with undifferentiated quantum
properties form non-local UDIs like in generative-GRW and in the
generative-quasi-local-MWI. More concretely, local interactions between the generators
and A/B lead A/B to have a stably differentiated quantum property (e.g.,
spin in a certain direction) and lead to a non-local stable
differentiation of a quantum property of B/A. This changes the position
determinate value of \(E_{B}/E_{A}\) when it interacts with B/A. Let's
suppose that a magnetic field acts on particle \(S_{A}\) in such a way
that \(E_{A}\) interacts with \(A\) where this interaction ends up
stably differentiating the spin-z of \(A\), changing the determinate
value of \(E_{A}\). Then, this also leads to the non-local stable
differentiation of the spin-z of \(B\). Furthermore, when \(E_{B}\) and
B interact, \(E_{B}\) will have a determinate value of position
influenced by the determinate value of B. Updating the state to the one
that resulted from the interactions, we end up, for example, with the
following quantum state,

\begin{equation}
|\Psi(t')\rangle = |\uparrow_{z}\rangle_{A} |\downarrow_{z}\rangle_{B} |\text{up}\rangle_{E_{A \text{ DS}}} |\text{down}\rangle_{E_{\text{B DS}}}
\end{equation}

Note that, like in other generative quantum theories, particles with indeterminate value properties play the role of the wave function in generative Bohmian mechanics. Generative Bohmian mechanics needs this dual ontology because, on any understanding, Bohmian mechanics has a dual ontology. So, this version of Bohmian mechanics is no more ``unnatural" than any other alternative version of this view.

\subsection{The generative-EnDQT and
generative-hybrid-classical-quantum-theories}
GQT was inspired by Environmental Determinacy-based Quantum Theory  (EnDQT) \cite{Pipa2023ATheory}. EnDQT is a local non-relationalist, non-superdeterministic, and non-retrocausal QT that makes indeterminacy basic. Besides being local (more on what I mean by local below), another benefit of EnDQT is that it is a conservative view since it doesn't modify the fundamental equations of QT. In this section, I will show how GQT helps generate EnDQT and how EnDQT provides a way to move beyond the MWI/GRW/Bohm orthodoxy altogether. In doing that, it also allows us to formulate a local interpretation of QT in the domain where
we know where to apply QT. I will also briefly go over hybrid classical-quantum theories, expressed via the lens of GQT, and explain some of their similarities with EnDQT.

Like the MWI and other relationalist views, for generative-EnDQT
(henceforward, EnDQT) particles don't play a fundamental
role. The stability conditions, i.e., the conditions under which a
system comes to have a determinate value/stably differentiated quantum property, can be
understood via four conditions that form the core of EnDQT. Plus, EnDQT
involves two hypotheses. Viewing EnDQT via GQT, we see that a key innovation of EnDQT is the
determination capacity (DC), which is the capacity that systems have to
give rise to other systems having their quantum properties stably
differentiated and to transmit the DC to other systems under
interactions. No other QT introduces this capacity, which is transmitted between systems under certain rules. As we will see, another innovation is the possible introduction of
a new kind of generator.

I will now go over the conservative determination
conditions (CDCs), which are the stability conditions for EnDQT in the non-relativistic domain.\footnote{More on this in \cite{Pipa2023ATheory} and further below.} They are called this way because I think that they are the most
conservative conditions for the DC to spread in this domain:\\

CDC1) The determination capacity (DC) of system $X$ concerning system $Y$
(DC-Y) is the capacity that $X$ has while interacting with $Y$,\\

\noindent i) to allow $Y$ to have a determinate value under the interaction with $X$ that also leads $X$ to have a determinate value, where $X$ and $Y$ have a determinate value in the same spacetime region,\vspace{5mm}

\noindent ii) to provide the DC to $Y$ concerning another system $Z$ (DC-Z) if and only if a) $Z$ starts interacting locally with $Y$ while $Y$ is already interacting with $X$, and b) $Y$ has a determinate value due to $X$ and $Z$ doesn't disturb this process.\\



So, the DC propagates between systems via interactions because $Z$ can
then have the DC concerning a system $K$ (DC-K), if and only a) $K$ starts
interacting with $Z$ while $Z$ is already interacting with $Y$, and b) $Z$ has a
determinate value due to $Y$, and so on for a system $L$ that interacts with
$K$ while $K$ interacts with $Z$, etc. Note that $X$ having a determinate value
and $Y$ having a determinate value in the interaction in i) is the same
event (i.e., it occurs in the same spacetime region). This is why, for
EnDQT interactions give rise to determinate values. 


As we have seen, the DC propagates between
systems via local interactions over spacetime, so interactions only
concern the regions where the systems are, where following the
standard way,\\

\noindent For a system $X$ to interact with system $Y$ from time t to t', the quantum
states of $X$ and $Y$ must at least evolve from t to t' under the
Hamiltonian of interaction representing the local interaction between $X$
and $Y$.\\

The chains of interactions that give rise to systems having determinate values and may propagate the DC are the so-called Stable
Determination Chains (SDCs). They are the DSs of EnDQT. All systems
that don't belong to SDCs belong to ISs. Furthermore, contrary to some
other QTs, there aren't interactions at a distance between systems that
compose the DSs or the ISs. Also, in agreement with GQT we have that\\

CDC2) Interactions between system $X$ and a set of systems that form a system $Y$ that may be larger, which has the DC, lead system $X$ to have a certain
determinate value, which corresponds to a certain quantum property
stably differentiated, where the distinguishability of the physical
state of $Y$ concerning the possible determinate values of $X$ allows us to
infer if $X$ will have a determinate value among the possible ones and
when that happens. Such distinguishability is inferred via the
decoherence of $X$ by $Y$, and where it is indeterministic, the values that
will arise among the possible ones.\\

Now, we can use CDC2) to spell out CDC1) in terms of decoherence (more
on how to understand decoherence according to EnDQT below).\\

CDC1*) The DC-Y of $X$ is the capacity that $X$ has while interacting with
Y,\\

i*) to decohere $Y$, which leads both systems to have a determinate value.
Let's suppose that system $S$ in eq. (\ref{decoherence}) is an instance of $X$, and system E
is an instance $Y$. The possible values of $X$ are represented by
\(\uparrow_{z}\) and \(\downarrow_{z}\). The possible values of $Y$ are
represented by \(E_{\uparrow}\) and \(E_{\downarrow}\).\\

ii*) to provide the DC-Z to $Y$ if and only if ii*-a) $Z$ starts interacting
with $Y$ while $Y$ is already interacting with $X$ and ii*-b) $Y$ is decohered by $X$, and $Z$ doesn't disturb this process, i.e., driving away to other states, the states of $Y$ that are being decohered by $X$.\\

CDC3) I will consider that two kinds of systems constitute an SDC:\\

\noindent -Initiator systems or initiators, which are systems that either a) have the DC concerning any system by default (i.e., they always have the DC-X for any system $X$), i.e., independently of their interactions with other systems. Or b) they are the first systems that have the DC concerning some system that they are interacting with or the ones that we initially assign in our models the DC concerning some system that they are interacting with. Because of this, initiators are the systems that start SDCs.\vspace{5mm}

\noindent -Non-initiator systems are systems that don't have the DC concerning a
system by default but have it due to their interactions with other
systems that have the DC.\\

So, for EnDQT, like in some other generative quantum theories, the world is
fundamentally constituted by systems with indeterminate
values/undifferentiated quantum properties, which include
initiators. The latter $may$ have the DC concerning any system by
default, not having to have their quantum properties stably
differentiated in a previous interaction to stably differentiate the
quantum properties of other systems and transmit the DC. On the other
hand, non-initiators have to have certain quantum properties stably
differentiated due to some previous interactions to have the DC. The DSs for EnDQT are called SDCs because the process that gives rise to them can be seen as a process that, in order to occur, needs to be stable in the sense that it stably obeys CDC1*). More concretely, in order to infer that systems have determinate values, it is necessary that the overlap terms of the quantum states of the environment go $stably$ to zero. Also, given CDC1*) again, it is also necessary that systems that interact with systems that are going over this process, are $stably$ not disturbed from going over this process.

Since systems are typically composed of many systems, EnDQT also assumes
that\\

CDC4) For a system $S$ to have the DC concerning some system $S'$, its
subsystems must have the DC concerning $S'$ or its subsystems.\\

Let's review the simple and idealized example.\footnote{See Appendix A in \cite{Pipa2023ATheory} for a toy model.} I will soon make this
example more concrete further below. I will again assume
that systems interact quickly compared with how quickly they
intrinsically evolve, so that, once again, we can neglect the systems'
intrinsic evolution. This example will involve systems \(S_{0}\),
\(S_{1}\), and \(S_{2}\), where \(S_{0}\) is an initiator, in a toy mini
universe where the SDC that will be formed has the following structure:
\(S_{0} \rightarrow S_{1} \rightarrow S_{2}\). The arrows represent the
stable differentiation of a quantum property of \(S_{1}\) by \(S_{0}\),
which allows \(S_{1}\) to stably differentiate a quantum property of
\(S_{2}\) and have the DC-\(S_{2}\).

Let's suppose that $S_0$  is an initiator of the kind a), and thus it has the DC concerning any system. Let's assume that \(S_{2}\) starts interacting with \(S_{1}\) while
\(S_{1}\) is already interacting with \(S_{0}\) so that \(S_{1}\) has the
DC-\(S_{2}\), and \(S_{1}\) can end up transmitting the DC to \(S_{2}\)
concerning some other system that \(S_{2}\) might end up interacting
with. However, in order to fulfill CDC1-ii) and to simplify, let's assume that when \(S_{1}\) and \(S_{2}\) begin interacting, the changes of the states of \(S_{1}\) that the interaction with \(S_{2}\) leads to is negligible in such a way that we can neglect the evolution of the quantum states of
\(S_{1}\) while \(S_{0}\) and \(S_{1}\) interact. Then, we can
idealize that \(S_{1}\) and \(S_{2}\) start interacting only when the
interaction between \(S_{0}\) and \(S_{1}\) ends.\footnote{We could
  similarly consider that while \(S_{0}\) interacts with \(S_{1}\),
  \(S_{2}\) starts interacting with \(S_{1}\) in such a way that it
  doesn't drive the states of \(S_{1}\) out of being states that
  \(S_{0}\) decoheres in the following sense: the Hamiltonian of
  interaction of \(S_{0}\) and \(S_{1}\) would still at least
  approximately commute with the (pointer) observable that these states
  are eigenstates of.} Thus, we can just analyze the evolution of the
quantum states of \(S_{0}\) while \(S_{0}\) and \(S_{1}\) are
interacting, where this interaction ends approximately at \(t'\), and
these systems have a determinate value at \(t'\).

Let's then put a subscript SDC on the quantum states of a system if that
system is an initiator or has the DC relative to some system. We then have the following interaction between \(S_{0}\) and
\(S_{1}\),

\begin{equation}
\begin{aligned}
    |E_{\text{ready}}\rangle_{S_{0} \text{ SD} C} (\alpha' |E_{0}'\rangle_{S_{1}} + \beta' |E_{1}'\rangle_{S_{1}}) \rightarrow_{t'} \\
    |E_{0}(t')\rangle_{S_{0} \text{ SD} C} |E_{0}'\rangle_{S_{1}} + |E_{1}(t')\rangle_{S_{0} \text{ SD} C} |E_{1}'\rangle_{S_{1}}
\end{aligned}
\end{equation}

So, if
\({{\langle E}_{0}(t')|E}_{1}(t')\rangle_{S_{0}\text{\ SD}C^{\ }}\approx0\)
and
\({{\langle E}_{1}(t')|E}_{0}{(t')\rangle_{S_{0}\text{\ SD}C^{\ }}}_{\ }\approx0\ \)quasi-irreversibly
when \(S_{0}\) and \(S_{1}\) end their interaction, \(S_{1}\) will have
a quantum property stably differentiated by \(S_{0}\) and a determinate
value of the associated quantum property (let's suppose that is either 0
or 1) that arises from its interaction with \(S_{0}\), and acquires the
DC-\(S_{2}\) (given our idealization). I am assuming that occurs at
\(t^{'}.\) Let's further assume that \(S_{1}\) has a determinate value
0. Then, the stably differentiated quantum property will be represented
by \({|E}_{0}^{'}\rangle_{S_{1}}\) and the observable that
\({|E}_{0}^{'}\rangle_{S_{1}}\) is an eigenstate of. Now, let's consider
the interaction between \(S_{1}\) and \(S_{2}\), which (assuming our
idealization) starts when the interaction between \(S_{0}\) and
\(S_{1}\) ends. Let's assume that it ends at \(t^{''}\),\footnote{Note
  that the quantum states of \(S_{1}\) and \(S_{2}\) absorbed their
  quantum amplitudes.}

\begin{equation}
\begin{aligned}
    |E_{0}'\rangle_{S_{1} \text{ SDC}} \left( \alpha |\uparrow\rangle_{S_{2}} + \beta |\downarrow\rangle_{S_{2}} \right) \rightarrow_{t''} \\
    |E_{0}^{' \uparrow}(t'')\rangle_{S_{1} \text{ SDC}} |\uparrow\rangle_{S_{2}} + |E_{0}^{' \downarrow}(t'')\rangle_{S_{1} \text{ SDC}} |\downarrow\rangle_{S_{2}}
\end{aligned}
\end{equation}

The evolution of the interaction between \(S_{1}\) and \(S_{2}\) can
be analyzed via the reduced density operator
\(\rho_{S_{2}}\left( t \right)\). This interaction will lead to the
stable differentiation of a quantum property of \(S_{2}\) and allow it
to have a determinate value (\(\uparrow\) or \(\downarrow\)) if
\(\langle E_{0}^{' \uparrow}\left( t \right)|E_{0}^{' \downarrow}\left( t \right)\rangle_{S_{1}} \approx 0\)
and
\(\langle E_{0}^{' \downarrow}\left( t \right)|E_{0}^{' \uparrow}\left( t \right)\rangle_{S_{1}} \approx 0\ \)quasi-irreversibly
when \(S_{1}\) and \(S_{2}\) end their interaction. Let's assume that
this interaction ends at \(t''\), and these systems will have a
determinate value at \(t^{''}\text{.\ }\)So, \(S_{1}\) will have indeterministically another
stably differentiated quantum property and a determinate value at
\(t''\) that arises from its interaction with \(S_{2}\), where the
possible values that it can have are represented via the eigenvalues of
the observable that
\({|E}_{0}^{' \uparrow}\left( t'' \right)\rangle_{S_2}\) and
\({|E}_{0}^{' \downarrow}\left( t'' \right)\rangle_{S_2}\) are eigenstates
of. Furthermore, \(S_{2}\) could have the DC concerning some other
system \(S_{3}\) if it interacted with it before its interaction with
\(S_{2}\) ends. In Appendix A in \cite{Pipa2023ATheory}, I have presented a more detailed toy model.

In the example above, $S_1$ could be a system well approximated by a large set of quantum harmonic oscillators\footnote{Or more precisely, bosonic modes. These interactions can be represented via the spin-boson mode. See, e.g., \cite{Leggett1987DynamicsSystem, Schlosshauer2007DecoherenceTransition}.} or a set of spin-1/2 systems or two-level systems that can be approximated as spin-1/2 systems. Each set of quantum systems interacts with a single spin-1/2 system. The collection of these systems constitutes system $S_2$. Then, system $C$ could be another two-level system that will interact with $S_2$.\footnote{The interactions between the systems that constitute $S_2$ and system $S_3$ would be modeled via the so-called spin-spin decoherence models developed in \cite{Zurek1982Environment-inducedRules, Cucchietti2005DecoherenceEnvironments}.}

As we can see, for EnDQT, irreversible decoherence is viewed as an inferential tool
that represents how the systems that are part of the nodes of SDCs
interact, and like generative-MWI and generative-Bohmian mechanics, to
infer the time it takes for stable differentiation to occur. However, EnDQT is an indeterministic theory, contrary to these theories. Also,
it is important to emphasize that now it is required that the systems
that belong to the environment have the DC, in order for determinate
values to arise. So, when there are interactions, but the systems
involved don't belong to an SDC, not having the DC, their relevant
quantum properties will remain undifferentiated. Thus, no determinate
value arises, and we don't update the quantum state to the new state.

Let's now review two hypotheses assumed by EnDQT, starting with the
one regarding decoherence. EnDQT has a subtler view of decoherence than
other QTs. Let's call the models of decoherence that represent the
interactions between systems having the DC, starting with the
initiators, \emph{fundamental decoherence models}. These models don't
involve extra considerations, such as if the environment is inaccessible
or open. The systems in CDC1-CDC4), and in the example above are
represented via these models.

On the other hand, the so-called
\emph{pragmatic decoherence models} don't necessarily track the
interactions with systems that have the DC. These models come in two kinds. We have seen them in the previous
sections, but I will distinguish them here again to precisify some of
their aspects and distinguish them from other processes.
\emph{Irreversible pragmatic decoherence models} are models that
represent situations where it is considered that is impossible to reverse
the process represented by them because they involve many systems, and
where these situations may involve an environment that is open. These
are the models typically associated with decoherence, which is a quasi-irreversible process. We also have what
I will call \emph{reversible pragmatic decoherence models}. These are
models that represent a process that apparently involves decoherence in
the sense that it is modeled by the overlap terms of the environment
going quasi-irreversibly to zero. However, someone in some privileged
position could reverse this process via operations on the systems or (to
put it less pragmatically) they don't involve enough degrees of freedom
to be considered irreversible. So, these models aren't what we associate
with decoherence.

Given the distinctions above, EnDQT also postulates the following
hypothesis regarding the structure of the SDCs:\\

\noindent The SDCs in our world are widespread in such a way that the empirically successful and local pragmatic irreversible decoherence models in open environments track the interactions between systems that belong to SDCs that serve as an environment for a target system that doesn't belong to an SDC, but ends up belonging to it. However, the SDCs in our world are such that there can also exist processes represented via local empirically successful reversible decoherence pragmatic models, where the latter are tracking the interactions between systems that don't belong to SDCs (\textit{SDCs-decoherence hypothesis}).\\

Via this hypothesis, EnDQT grounds the success of these pragmatic
decoherence models in representing processes that give rise to
determinate values. It is important to notice that depending on one's
ingenuity, \emph{in principle,} it is possible to isolate macroscopic
systems from the influence of SDCs, and so for EnDQT, in principle,
arbitrary systems can be in a coherent superposition for an arbitrary amount of
time. Thus, if this isolation is done properly in such a way that we can
unitarily manipulate the contents of that region, we might have a
process of reversible decoherence inside that region instead of an
irreversible one. So, given the above hypothesis, if some situation,
even involving interactions between macroscopic systems, is
appropriately modeled by reversible pragmatic decoherence models, we can
infer that we have managed to isolate the systems from the influence of
SDCs. Of course, also given this hypothesis, in principle, doesn't mean
in practice because our pragmatic models of decoherence tell us that
it is very difficult to place large macroscopic systems in a
superposition.

This view held by EnDQT contrasts with the one often assumed by MWI-like
views, which would consider that determinacy arises within a large
enough isolated spatiotemporal region with systems decohering each other
inside of it. In the case of the MWI, the DC doesn't exist and matter. Note again that, contrary
to most of the previous QTs, for EnDQT there aren't any
non-local ISs or DSs connecting systems. Those structures arise and are
maintained locally via their interactions. SDCs for EnDQT can be
represented by directed graphs like the one in the example above, where
the arrows represent the stable differentiation interactions arising
between systems.

Now, we are in a better position to further clarify how EnDQT relates to some of
the other features of GQT. Contrary to the other QTs explained here
(more on this in the next section), EnDQT has the benefit of explaining
in a unificatory and parsimonious way\footnote{Note that initiators of the kind a) don't lead so much to a parsimonious theory, but the ones of the kind b) because they are identical to non-initiators.} via the initiators and the laws
that describe/govern the interactions of systems that belong to the SDCs
they give rise to, which systems can be generators and the generative
quantum properties that they have. More concretely, initiators are a
special kind of generators that have the capacity of allowing other
systems to become generators when they interact with them. The
generative quantum properties of generators are the properties that they
have when they interact with other systems, giving rise to the latter
having determinate values. Above, ${|E}_{0}{\left( t' \right)\rangle}_{S_{0}\text{SD}C} \quad \text{or} \quad {|E}_{1}{\left( t' \right)\rangle}_{S_{0}\text{SD}C}$
and the respective observable that these quantum states are eigenstates of, represent those generative quantum properties. Systems that they interact with will be able to have determinate values and certain generative quantum properties, becoming generators. The possible generative quantum properties of \(S_{1}\) are represented by
${|E}_{0}^{' \uparrow}(t'')\rangle_{S_{1}} \quad \text{or} \quad {|E}_{0}^{' \downarrow}\left( t'' \right)\rangle_{S_{1}}$, and the observable that these quantum states are eigenstates of, which gives rise to \(S_{2}\) having a determinate value. So, we can trace the capacity of systems having generative quantum properties and being
generators to interactions that ultimately originated with initiators.

The second hypothesis aims to address the question of what kind
of systems initiators are. The inflaton is one possible candidate for an
initiator because of its privileged and influential role in the history
of the universe, which accounts for our belief that systems with
determinate values are widespread (i.e., classicality is
widespread). So, the inflaton field, with its
quantum properties occupying regions of spacetime, would be the
initiator.\footnote{To clarify, notice that ultimately, the description of the inflaton field and the rest of the fields interacting with it would need to be
quantum field theoretic. I haven't shown how GQT can be understood in
the context of quantum field theory, but in principle, such an extension
won't be problematic. Briefly, in one possible approach, the fundamental systems that transmit the DC are quantum fields in a finite spacetime region developing local interactions. More concretely, we associate to quantum fields in a finite spacetime region, such as the inflaton field $\phi$ in the spatial region \(x\) at t (whose observable
is \(\hat{\phi}\left( x,\ t \right)\)), collections of quantum
properties represented via the wave functional
\(\Psi\lbrack\phi,\ t\rbrack\) (see, e.g., \cite{sep-quantum-field-theory} for an introduction
  to quantum fields). The wave functional assigns a complex amplitude to each
possible configuration of classical fields in that 
 spacetime region,
yielding a superposition of these configurations. The quantum properties
of the quantum system (which concerns that region) will also be
represented by the observables that act on the wave functional in that
region. So, although quantum fields, in general, can be associated with unbounded
spacetime regions, the DC is transmitted between localized quantum fields
via local interactions since the observables
(including the field observables such as
\(\hat{\phi}\left( x,\ t \right)\)) representing quantum systems and quantum
properties concerns bounded spacetime regions developing local interactions. Thus, it is expected that
quantum fields in certain regions will have stably differentiated quantum properties and
spread the DC in local regions of spacetime.} The inflaton field is the
initiator, but it transmits the DC via local interactions.

 If we adopt initiators of the kind a), one of the reasons to consider the inflaton field (and the quantum
systems that arise from this field) as a plausible candidate for an
initiator is that it allows us to explain why we can sometimes maintain
the coherence of quantum systems in quasi-isolated spatiotemporal
regions. If there were initiators that could start SDCs in any region,
it would be very difficult or impossible to maintain such coherence
because they would destroy superpositions. We can allow for initiators
that only manifested themselves at the beginning of the universe by
observing that it is standardly considered that, at least in our
universe, the inflaton field reached the absolute minimum of its
potential and has been staying there.\footnote{See, e.g., \cite{Liddle2009TheStructure}.} Then, for example, if we consider the condition that this
minimum corresponds to the point where the field is zero and if we
consider that the coupling of the inflaton field to all other fields in
the Lagrangian density that describes/governs our universe depends on
the value of the inflaton field in such a way that the interaction terms
representing these interactions are zero when the field zero, we can
consider that the inflaton field in the stages of the evolution of the
universe after the reheating phase will at least rarely interact with
other fields/systems.\footnote{See, e.g., \cite{Kiefer:2008kub} for
  models of decoherence involving the decoherence of the inflaton field,
  and a discussion of the various possible kinds of environmental
  systems. See also \cite{Pipa2023ATheory}.} So, it will (at least) rarely give rise
to SDCs after the reheating phase, which is our current phase. Let's
represent the Lagrangian of our universe obeying these conditions as
\(\mathcal{L}_{\text{SDC}}\). So, the second hypothesis is that\\

\noindent at least most current SDCs started in the early universe, and initiators
had a privileged role in this stage, giving rise to these SDCs, and
where the initiators are the inflaton field described via
\(\mathcal{L}_{\text{SDC}}\) (inflationary-starting hypothesis).\\

This is one possible concrete hypothesis for what initiators are, being
an instance of the more abstract \emph{SDCs-starting hypothesis}, which
establishes when SDCs started.  I regard this
latter hypothesis as a placeholder for current and future cosmology.
Given the current evidence for inflation, this is the initiator adopted.
As I have argued in \cite{Pipa2023ATheory}, the possible specialness of initiators is, in
principle, unproblematic because our evidence points toward early
universe events involving some special physical phenomena and can
provide other scientific and philosophical advantages (more on this in
the next section).

An important benefit of EnDQT is that it deals with Bell's theorem and scenarios in the sense of not leading to the violation of relativistic causality, i.e., without forcing us to assume that the
causes of the events involved in those correlations aren't in their past
lightcone and without invoking superdeterministic or retrocausal
explanations, being ``local''.\footnote{When I say that the causes are in the ``past
  lightcone,'' I am implicitly assuming that this influence is
  non-instantaneous.} First of all, note that EnDQT doesn't modify the
fundamental equations of QT, and so, in principle, it can be rendered
Lorentz and generally covariant, and thus, it can be compatible with relativity in
this sense as long as QT is.\footnote{Within the domain where we know where to apply QT. See \cite{Pipa2023ATheory}.}
However, future work will provide a model of EnDQT where this is shown
explicitly. Second, let's see how it deals with the EPR-Bell scenarios
and Bell's theorem.


A widely accepted version of Bell's theorem involves, together with the statistical independence or
no-superdeterminism assumption,\footnote{This assumption states that any
  events on a space-like hypersurface are uncorrelated with any set of
  interventions subsequent to it.} the factorizability condition,

\begin{equation}
P(AB|XY\mathbf{\Lambda}) = P(A|X\Lambda)P(B|Y\mathbf{\Lambda})
\end{equation}

\begin{figure}[h!]
    \centering
    \includegraphics[width=\linewidth]{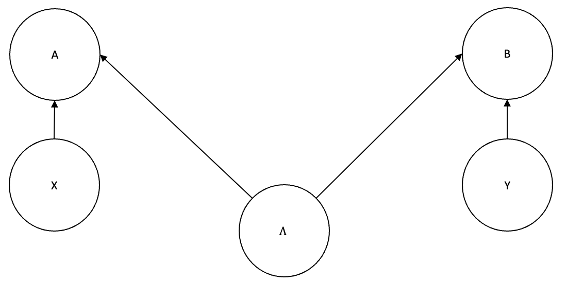}
    \caption{DAG of the common cause structure of Bell correlations, which respects relativity. This causal structure respects relativistic causality because \(X\) or \(A\) doesn't influence \(Y\) or \(B\), and vice-versa, where these events may be spacelike separated. Moreover, no other variables influence the variables \(A\), \(B\), \(X\), or \(Y\), or they don’t influence anything else. So, there are no retrocausal or superdeterministic causal relations.}
    \label{DAGcausal}
\end{figure}

The variables $A$, $B$, \(\mathbf{\Lambda}\), $X$, and $Y$ concern events embedded in a Minkowski spacetime. $A$ and $B$ represent the different measurement results of Alice and Bob, $X$ and $Y$ are the different possible choices of measurement settings for Alice and Bob. \(\mathbf{\Lambda}\) represents some set of (classical) ``hidden'' variables in the past lightcone of $A$ and $B$ (see also Figure \ref{DAGcausal}), representing the common causes of the correlations between $X$ and $Y$. This condition is seen as a consequence of two assumptions:\footnote{\cite{Bell1976TheBeables, Bell2004LaCuisine}. See also, e.g., \cite{sep-bell-theorem} and references therein.} the causes of an event are in its past lightcone, and the classical Reichenbach Common Cause Principle (CRCCP).

Briefly, the CRCCP states that if events $A$ and $B$ are correlated, then
either $A$ causes $B$, or $B$ causes $A$, or both $A$ and $B$ have common causes
\(\mathbf{\Lambda}\), where conditioning on the members of the set of variables \(\mathbf{\Lambda}\), $A$ and
B are decorrelated, i.e.,
\(P(A,\ B|\mathbf{\ \Lambda)} = \mathbf{\ }P(A|\mathbf{\ \Lambda)}P(B|\mathbf{\ \Lambda)}\).
However, it is unclear whether we should accept that the probabilistic
relations and conditions given by the CRCCP should, in general,
represent a causal structure involving quantum systems, given their
quantum indeterminate values, and how they evolve. The CRCCP. However, why should we trust the CRCCP as a general statement about causal relations? The CRCCP as a general statement is better seen
as a consequence of the more widely applicable Classical Markov Condition (CMC), assumed by the widely applicable
Classical Causal Models (CCMs).\footnote{I will not derive it here, but
  see \cite{sep-physics-Rpcc}. See \cite{Pipa2023ATheory} for some subtelities regarding this relation.}

The CMC connects the causal structure provided by some theory
represented by a DAG with probabilistic statements. The CMC is the
following,\\

\noindent Let's assume we have a DAG \(G\), representing a causal structure over the variables \(\mathbf{V} = \{ X_{1},\ \ldots,\ X_{n}\}\). A joint probability distribution \(P\left( X_{1},\ \ldots,\ X_{n} \right)\) is \emph{classical Markov} with respect to \(G\) if and only if it satisfies the following condition: for all distinct variables in \(\mathbf{V}\), \(P\) over these variables factorizes as $P\left( X_{1},\ \ldots,\ X_{n} \right) = \prod_{j}^{}{P\left( X_{j} \middle| \text{Pa}\left( X_{j} \right) \right)}$, 
where \(\text{Pa}\left( X_{j} \right)\) are the "parent nodes" of \(X_{j}\), i.e., the nodes whose arrows from these nodes point to \(X_{j}\).\\

The CMC for the DAG in Figure \ref{DAGcausal}, which respects relativity, allows us to
derive the following equation (I will denote certain regions of
spacetime, the related nodes, and variables whose values may be
instantiated in those regions using the same letters),

\begin{equation}
P(AB|XY) = \sum_{\Lambda} P(\Lambda) P(A|X\Lambda) P(B|Y\Lambda).
\end{equation}

The acceptability of the CRCCP can be supported by the empirical success
of the application of the CMC via CCMs (e.g., \cite{Pearl2009Causality}). EnDQT
responds to Bell's theorem by rejecting that the CMC can be applied in
general to accurately represent causal relations between quantum
systems, understood here simply as relations of influence. Hence, it
rejects the applicability of the CRCCP and the factorizability condition
to make such an accurate representation.

In \cite{Pipa2023ATheory}, I have provided various reasons why EnDQT rejects that
the CMC and CCMs accurately represent causal relations between quantum systems.


One of the reasons comes from noticing that a clear and precise way of justifying the CMC is via Pearl \&
Verma's proof \cite{Pearl1995ACausation} of the CMC. This proof roughly assumes that the
Markov condition arises from the ignorance of some absolutely defined
``hidden variables'' \(\Lambda\) that are the common causes of
correlations, where the latter represents some determinate values that
determine these correlations. However, EnDQT rejects that common causes of the correlations are  given by such
determinate values that we are ignorant about. As we shall see, they are appropriately given by
quantum states representing quantum indeterminate values.\footnote{The argument in \cite{Pipa2023ATheory} is more subtle, but I will simplify it here.}
Note that even if we specify some \(\Lambda\) via the eigenvalues of the
observable that some quantum state in the past is an eigenstate of,
given that EnDQT is an indeterministic theory, such \(\Lambda\) can't
work to specify a local common cause of Bell correlations. So,
such specification wouldn't be appropriate anyway to infer the relations
of influence behind Bell correlations. Thus, the CMC and the CCMs are inappropriate to accurately represent causal relations between quantum systems.\footnote{I reflect further on this argument in \cite{Pipa2023ATheory}.}

Further below, I will provide other related reasons why EnDQT rejects that the CMC and CCMs accurately represent causal relations between quantum systems.

Instead of the CMC, EnDQT uses a generalization of the
CMC, the quantum Markov condition (QMC), and Quantum Causal Models
(QCMs)\footnote{\cite{Costa_2016, Allen2017QuantumModels, Allen2017QuantumModels}.} that adopt a quantum version of the CMC as a more appropriate tool to accurately represent causal relations between quantum systems. As we will also see, QCMs provide a
local common cause explanation of Bell correlations. This will show how a kind of DSs (i.e., the SDCs) and ISs help provide that explanation (more on this in the next section).

This explanation is done via the quantum Markov condition and a version of the Born rule (Figure \ref{fig:collapseDS}),

\begin{equation}
P\left( x,y \middle| s,t \right) = \text{Tr}_{\Lambda \text{AB}} \left( \rho_{\Lambda} \rho_{A|\Lambda} \rho_{B|\Lambda} \tau_{A}^{x|s \text{ SDC}} \otimes \tau_{B}^{y|t \text{ SDC}} \right).
\end{equation}

Now, $A$, $B$, and $\Lambda$ represent spacetime regions instead of classical variables. The systems that are prepared at the source are acting as common causes
for Bell correlations. They have indeterminate values, until
non-instantaneously reaching the measurement devices of Alice and Bob,
which gives rise to the correlated outcomes. The entangled state
$\rho_{\Lambda}$, through its subsystems, represents these
systems that are prepared at a source, which, for instance, can be
systems having indeterminate values of spin-p, where p is ranging over
all possible directions of spin. \(\rho_{\Lambda}^{\ }\ \) and the
quantum channels \(\rho_{B|\Lambda}^{\ }\) and \(\rho_{A|\Lambda}^{\ }\)
are used to separately represent each system prepared at the source that
travels non-instantaneously to different regions, and this is done
by keeping track of the labels $A$ and B.\footnote{See \cite{10.5555/1972505}.} So, each one of the systems evolves locally to region $A$/$B$,
where Alice/Bob will influence the outcomes arising in those regions.
When it comes to A, this influence is represented via the quantum
channel \(\rho_{A|\Lambda}^{\ }\ \), and when it comes $B$, by
\(\rho_{B|\Lambda}^{\ }\). More concretely, \(\rho_{B|\Lambda}^{\ }\)
and \(\rho_{A|\Lambda}^{\ }\ \) are identity channels that via their
action on the density operator \(\rho_{\Lambda}^{\ }\ ,\ \)also
represent the systems in region $\Lambda$ that evolve to
regions $A$ and $B$, respectively. The influence that gives rise to the
outcomes is also represented via the POVMs \(\tau_{A}^{x|s\ SDC}\) in
Alice's case, where \(s\) is her random measurement choice, and x is her
outcome, and via \(\tau_{B}^{y|t\ SDC}\) in Bob's case. 

\begin{figure}[H] 
    \centering
    \includegraphics[width=0.8\textwidth]{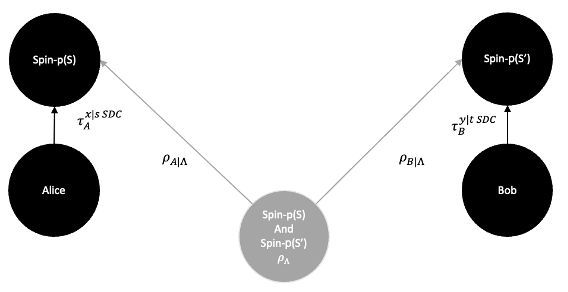} 
    \caption{\emph{EnDQT-causal-DAG}~proposed by EnDQT, which allows for a non-relational local common cause explanation of Bell correlations.}
    \label{DAG-EnDQT} 
\end{figure}

The superscript
SDC placed in the POVMs means that these are interventions/interactions
that give rise to a determinate value, involving systems that are part
of a local SDC, making others also part of an SDC. These interactions
are represented by other types of edges in the DAG in Figure \ref{DAG-EnDQT}. In this
case, Alice and Bob, due to their measurements, will lead the target
systems to be part of an SDC because they also belong to SDCs giving rise to the systems having a determinate value of spin
in a specific direction, for
example. So, via the above account, EnDQT uses QCMs to
give a local common cause explanation of quantum correlations, which
respects relativistic causality. It is important to notice that, by
assuming GQT's perspective on quantum states, which doesn't reify them, and by assuming local DSs (the SDCs),
we shouldn't consider that the measurement of Alice on her system
influences Bob's system and vice-versa.

This scenario can be represented using a DAG, which I have called
\emph{EnDQT-causal-DAG}. The diagram shows in grey the evolution of systems that
are not part of an SDC but rather an IS concerning the quantum
properties of the model. The evolution and interactions of
systems that belong to an SDC are in black:\footnote{See \cite{Pipa2023ATheory} to
  see how EnDQT can account for interference locally.}


We see above the role of structural generators in giving a local common
cause explanation of Bell correlations.

 In \cite{Pipa2023ATheory}, I have provided other reasons why EnDQT rejects that the CMC and CCMs accurately represent causal relations between quantum systems (see above some other related reasons). This argument, which is called the \textit{argument for locality}, is based on finding the domain of applicability of the CCMs by examining the more general models that putatively represent causal relations in the quantum domain, i.e., QCMs, which reduce classical ones in a ``classical limit.'' Like we found what is wrong with classical mechanics when we examined the more general theory, QT, which reduces to classical mechanics in some limit, we find what is wrong with the CCMs when we adopt QCMs interpreted via EnDQT. This argument is necessarily theory-dependent because it relies on an interpretation of CCMs and QCMs. Like all frameworks, these models always require an interpretation. The idea is that contrary to QCMs, CCMs with their CMC for EnDQT don't accurately represent relations of influence between quantum systems. This is because, contrary to CCMs, QCMs, as interpreted by EnDQT, explicitly consider that systems that are common causes for Bell correlations,\\
 
 i) only assume determinate values with a certain probability given by the Born rule when they interact  with systems that belong to SDCs. Therefore, they don't consider that there are certain hidden variables, which represent those determinate values and determine those outcomes. Furthermore,\\

 ii) the relations of influence for QCMs that give rise to quantum correlations are described via QT but without reifying the quantum states in the sense of EnDQT.\\

 Thus, given i) and ii), for EnDQT QCMs don't lead to inferences that consider that Alice affects the space-like separated Bob and vice-versa in EPR-Bell scenarios; Alice and Bob, in EPR-Bell scenarios, interact locally with their target systems via local SDCs (where each system is represented via \(\rho_{A|\Lambda}^{\ }\ \)), not influencing each other non-locally. Furthermore, no non-local ISs are assumed (or needed) like in the QTs above. i) and ii) allow EnDQT to interpret QCMs non-instrumentally and as clearly representing local features of the world, i.e., as not hiding non-local influences \emph{behind} quantum states and interventions on systems.

 Therefore,  contrary to QCMs, given that CCMs interpreted by EnDQT don't explicitly assume i) and ii), CCMs with their CMC for EnDQT don't accurately represent relations of influence between quantum systems behind quantum correlations.


It is sometimes seen in the literature the argument that QT is
non-local based on the claim that the EPR argument \cite{Einstein1935CanComplete} ruled out the existence of local indeterministic theories (e.g.,
\cite{Maudlin_2014}). So, one might worry that there is something wrong with
my argument above. However, this argument concerning the non-locality of
QT shouldn't be right because EnDQT, as an indeterministic local theory,
is a counterexample to that claim. Note that the so-called EPR
criterion of reality\footnote{``If, without in any way disturbing a
  system, we can predict with certainty (i.e., with probability equal to
  unity) the value of a physical quantity, then there exists an element
  of reality corresponding to that quantity.'' \cite{Einstein1935CanComplete}}
assumed in this argument can precisely be seen as a consequence of the
classical Reichenbach common cause principle (see \cite{Gomori2021OnCriterion}, which, as I have mentioned, is a special case of the more general
CMC \cite{sep-physics-Rpcc}). However, as I have argued in more detail in \cite{Pipa2023ATheory}, EnDQT doesn't consider that the
CMC can, in general, represent causal relations between quantum systems.
Thus, it rejects the EPR criterion as representing such causal relations
and, therefore, one of the basic premises of the above argument for
non-locality.

Another generative quantum theory leads to a version of hybrid classical-quantum theories,\footnote{See \cite{Oppenheim2023} and, e.g., \cite{Diosi1995QuantumLimit}.}
which I will call generative-hybrid. Due to reasons of space, here I
will go over it briefly. In the case of this theory, we have a kind of gravity-causes collapse theory, where there are classical systems that
evolve fundamentally stochastically,\footnote{So that the gravitational
  field doesn't ``reveal'' the location of the quantum systems in its
  interactions with them, \emph{collapsing} their quantum states, in
  certain situations in agreement with experiments. However, the greater
  the rate of \emph{decoherence} induced by classical systems on quantum
  systems, the lower the amount of diffusion/stochasticity induced by the
  quantum systems on the metric and their conjugate momenta of the
  classical system \cite{Oppenheim2023GravitationallyGravity}.} and quantum systems. The
evolution of both is described/governed by a hybrid classical-quantum
dynamics. Quantum systems, by default, belong to an IS and have
undifferentiated quantum properties like EnDQT, and like EnDQT, there
aren't any non-local ISs. Furthermore, classical systems are a
collection of quantum properties that are always stably differentiated,
e.g., the metric and its conjugate in the hybrid theory that aims to
describe gravity, and occupy spatiotemporal regions.\footnote{I don't
  regard calling these properties quantum or not as a substantive issue
  in a fundamental theory. Like in GRW, a quantum property doesn't need
  to be represented via standard QT.} Note that the metric as a field
will have values throughout all spacetime, and like in the case of
quantum systems when we have quantum fields,\footnote{See footnote 48.}
classical systems will pertain to certain spacetime regions.
The stochastic behavior of the metric is represented via a positive
density operator. A classical-quantum state is the tensor product
between these operators and quantum states/density operators. A
classical-quantum system is a collection of the quantum properties of
both, concerning certain regions of spacetime. Classical systems are
generators and stably differentiate the quantum properties of quantum
systems, and the latter also backreacts on the classical systems,
affecting their evolution. DSs concern the local evolution of these
classical systems and their interactions with quantum ones.
Generative-hybrid is local and can also provide a local causal
explanation of Bell correlations like EnDQT (more on this
below),\footnote{This is why hybrid classical-quantum theories shouldn't
  reify the quantum state because this doesn't lead to inferences about
  non-local causal relations, adopting GQT point of view.} but gravity
is necessarily the sole responsible for determinate values arising.

\section{GQT vs. Wave Function Realism and
Primitive Ontology and new generative quantum theories} 

Let's now compare GQT with Wave function Realism (WR) and Primitive Ontology (PO). I will argue that GQT has important benefits that
these frameworks don't offer and without some of their notorious costs.



First, GQT offers a better way of making sense of the nature of the
wave function or quantum state or density operators or matrices than WR
and PO (I will refer to density operators also as quantum states from
now on).\footnote{I am also assuming that PO proponents can assume that
  density operators have a nomological character.} WR is an ontology
that considers that the fundamental entity represented by QT is a
wave function living in a $3N$ configuration space where $N$ is the number of
existing particles.\footnote{See Section 1 for further information on WR.} As it is
well-known, the main challenge of WR is to give a plausible account of
how to derive and make sense of the spatial three-dimensional manifest
image from this more fundamental space. This is problematic, given the
evidence that we have that, at least in the classical regime, systems
occupy regions of spacetime.
This brings me to the PO. According to this view, what is fundamental
are entities with determinate locations in spacetime, having determinate
features, like flashes, mass densities, etc., also known as local
beables. This contrasts with the GQT since, in the latter case,
fundamental entities can have indeterminate locations. Another feature
of the PO framework is their view of wave functions/quantum states,
which are to be considered fundamental but it is typically considered
not to represent matter. It typically rather has a nomological
character, governing or describing the behavior of the PO. Although PO advocates may also allow the wave function to
be a physical wave in a high-dimensional space, I will specialize my
discussion on the former case since it sets it more apart from GQT and
WR. PO endorses a revisionist attitude towards the laws of nature by
considering that a complicated object, such as a wave function/quantum
state, which is also allowed to change over time, is a law.\footnote{One
  strategy tries to address this worry by arguing that the fundamental
  wave function of the universe behaves more like a law and may be
  simpler because it doesn't change over time \cite{Goldstein201391RealityTheory}. This is supported by the Wheeler-DeWitt equation assumed by
  some theories of quantum gravity. However, this strategy is highly
  speculative because it relies on an assumption that not all quantum
  gravity programs make.}

GQT doesn't suffer from the issues associated with considering the
wave function as an entity in a 3N dimensional space or a law.
The quantum state is more like a distribution over a set of
possibilities, and possibilities, unlike laws, change over time.
Furthermore, although systems may have certain indeterminate locations,
they can still occupy regions of spacetime. Thus, GQT offers a better
way of making sense of the nature of quantum states or the wave function
than WR and PO.

Second, contrary to these other frameworks, GQT is built in such a way
that helps formulate new QTs, which may lead to scientific and
philosophical progress. For example, GQT, in principle, can help
generate new QTs (and associated ontologies) that adopt a
strategy compatible with relativistic causality, i.e., that are local in the sense of not requiring us to assume that the causes of the events involved in those correlations are not within their past lightcone, and without invoking superdeterministic or retrocausal explanations. 
Given the importance of relativistic causality, this should be regarded
as a benefit. As I will argue, PO and WR lack this benefit.

We have seen EnDQT and a local version of the MWI above as examples of
theories that GQT helped generate. Furthermore, by changing some of the
features that EnDQT adopts, GQT can, in principle, help generate
theories that are a hybrid of EnDQT and other QTs. As I
will argue, these hybrids lead to other theories that allow
for the compatibility between QT and relativistic causality, plus have
some other benefits. Note that here I am not interested in locality in the sense that these theories are compatible with relativistic symmetries. This is an open question.\footnote{Note that, however, in the case of EnDQT-MWI and other theories that don't modify the fundamental equations of QT that we will see below, in principle, they allow for general covariance like EnDQT in principle allows. This is because they don't modify the fundamental equation of QT, and if QT allows for it (which is a conjecture), they also should allow for it. See Section 2.4.} I will sketch three examples of these theories here and leave their development for future work.


Regarding what I will call \emph{EnDQT-collapse}, it adopts a
different kind of generators and, more specifically, initiators than the ones adopted by EnDQT. Systems
with a specific quantum property (e.g., the position quantum property)
can become indeterministically initiators with a probability per unit
time and start an SDC like in EnDQT. Regarding what I will call
\emph{EnDQT-MWI}, it arises from considering a theory like EnDQT, but
where the generators of EnDQT don't give rise to determinate values
indeterministically in a single world but instead give rise to multiple
systems with determinate values deterministically, where each
corresponds to a world. We could perhaps also have an
\emph{EnDQT-Hybrid} that arises from modifying the generative-Hybrid by
introducing initiators in them. In EnDQT-Hybrid, for example, initiators
give rise to stochastically evolving classical systems with certain
quantum properties always stably differentiated. Or, we wouldn't have
initiators as systems but as the events that are behind the decoupling
between the classical properties (i.e., always stably differentiated
quantum properties that evolve stochastically in a certain way) and the
quantum ones. So, EnDQT-Hybrid would have the advantage of giving a
unifying explanation for why we have both classical and quantum systems.\footnote{A mechanism to describe these kinds of initiators should be explored in
future work.}

GQT helps generate versions of ``consciousness causes collapse theories,''\footnote{See, e.g., \cite{Wigner1961RemarksQuestion, Stapp1993MindMechanics, Chalmers2022ConsciousnessFunction}.} without the need to modify the equations of QT. Consciousness here is understood as phenomenal consciousness. A system is considered conscious when there is something that is like to be that system from within the system.\footnote{\cite{Chalmers2022ConsciousnessFunction}.} Maintaining the (non-mandatory) ontology of quantum properties, there are multiple versions of this idea. For instance, in one version, only generators with stably differentiated quantum properties are conscious systems. In another version, not only the generator but also its target system of an interaction has a stably differentiated quantum property. Furthermore, systems with stably differentiated quantum properties would be determinately conscious. Since stable differentiation comes in degrees, one can formulate a version where being conscious comes in degrees, where the degree of stable differentiation of a quantum property of a system tracks the degree to which a system is conscious (conscious of what? more on this below). When a system has an undifferentiated quantum property, it is indeterminately conscious (in one subversion), or it has no consciousness (in another subversion). 

However, let's focus on the first version that was mentioned, which is the most straightforward one. As I have mentioned, in this version, only generators with stably differentiated quantum properties are conscious systems. There are at least two generative versions of this view.  In the generative consciousness version, generators are systems that have the capacity to be conscious, and an indeterministic process gives rise to a generator being conscious. The Born rule gives the probabilities for this chancy process to occur.  Decoherence would be an epistemic and inferential tool that helps investigate when we have generators that become conscious, having stably differentiated quantum properties. This is when they $decohere$ other systems. In virtue of being conscious, systems give rise to others having stably differentiated quantum properties, where the values concerning these properties are absolute. Furthermore, they have experiences concerning such quantum properties. For instance, having the consciousness of energy would arise upon decoherence of some target system $S$ in the energy basis (stably differentiating this quantum property of $S$), having consciousness of momentum would arise upon decoherence of some system in the momentum basis, and so on for other quantum properties.\footnote{Here, we could adopt the view that it is the environment as a whole that is the generator, and it is the whole that is conscious, and none of its parts are. Or we could adopt the view that all the subsystems of the environment are generators, and all of them are conscious.} 

The problem with this view is that it inherits all the vagueness associated with decoherence. As I have explained in the previous sections, some processes apparently involve decoherence, but such decoherence is reversible and is not really considered decoherence. How to make that distinction? Unfortunately, for most quantum systems, we can't ask them whether they are conscious or not. One may argue that via the conditions for irreversible decoherence to occur, we would simultaneously investigate the conditions when consciousness arises, and we have multiple ways of finding what those conditions are. This often occurs when we have a large number of systems in an open environment. However, those conditions are still unclear because even what counts as an environment is still vague. The proponent of this view can appeal to a set of adequate theories of consciousness to inform them of what are the favorable conditions. This option would mean that generative consciousness is incomplete because it needs to appeal to other theories that very likely won't be physical theories. I think that this incompleteness renders this view less attractive.

One way of possibly surpassing this limitation is via an alternative version that I will call generative EnDQT-consciousness. This version is just like EnDQT (it has initiators, the DC, etc.), but it is considered that only systems with the DC are capable of being conscious. They become conscious when they have a stably differentiated quantum property and the DC. The capacity for a system to be conscious is transmitted via interactions. However, it is hard to see what this version adds to the original version of EnDQT. It rather seems to complicate it because now we have introduced consciousness at the fundamental level. Besides that, given our current science of consciousness, it is unclear whether stable differentiation of quantum properties of certain systems (or some analogous feature if we adopt some other ontology of properties) has anything to do with consciousness. Thus, it seems that we have introduced an unnecessary speculative assumption, and we might as well not assume it.

All theories mentioned above could use QCMs to provide a local common explanation of Bell correlations. Also, like EnDQT, they could formulate a version of the argument for locality explained in section 2.4 and in more detail in \cite{Pipa2023ATheory} to deal with the Bell's theorem in a local way, as well as other similar strategies to EnDQT to deal with this theorem mentioned in section 2.4. This is because, similarly to EnDQT, in principle, they consider that systems that are common causes for Bell scenarios i) only assume determinate values with a certain probability given by the Born rule when they interact (locally) with generators, where these local interactions are guaranteed by their local DSs, and ii) where those relations are described via QT but without reifying the quantum states like EnDQT assumes. Thus, given i) and ii), these theories don't reify the quantum states,
which doesn't lead to inferences that consider that Alice affects the
space-like separated Bob and vice-versa in EPR-Bell scenarios; Alice
and Bob, in EPR-Bell scenarios, interact locally with their target systems
via local DSs, not influencing each other non-locally; and no non-local ISs are assumed or needed like in the QTs
above. So, i) and ii) allow these theories to interpret QCMs
non-instrumentally and as clearly representing local features of the
world, i.e., as not hiding non-local influences \emph{behind} quantum
states and interventions on systems. Also, as I have mentioned above, all of them could, in principle, a version of the argument for locality, which uses i) and ii), to deal with Bell's theorem (see Section 2.4 and \cite{Pipa2023ATheory}), arguing that what it shows is that the Classical Markov Condition and the Classical Causal Models don't accurately represent causal relations between quantum systems. So, they can use quantum causal models to provide a common cause explanation of quantum correlations like EnDQT. Rather, Quantum Causal Models adequately represent such relations. However, I want to emphasize that instead of providing the argument for locality via SDCs, the argument would be given in terms of their own local DSs and the ISs represented via QCMs and other tools such as directed graphs. 

Furthermore, like EnDQT they can also reject the applicability of the Classical Markov Condition as accuratelly representing causal relations between quantum systems by rejecting Pearl and Verma justification of it. However, one from EnDQT that one should keep in mind is that local-MWI and EnDQT-MWI are deterministic and not indeterministic like EnDQT and the other views. Besides, they assume that there aren't single outcomes. So, these theories deal with
Pearl and Verma's proof of the Causal Markov Condition \cite{Pearl1995ACausation} by also rejecting that there are single outcomes of interactions, which is another assumption of that proof.

So, if quantum states concern systems with indeterminate values and the
possible determinate values, as well as DSs and ISs, whether a theory
obeys relativistic causality depends on the details of how the
indeterminate values become determinate via ISs and DSs. Thus, by allowing
for fundamental systems to have indeterminate value properties and not
reifying the wave function or seeing it only as nomic, we gain the
benefit of being able to add instead a different structure, which allows
for locality in certain QTs. For instance, we saw that EnDQT appeals to
ISs without non-local interactions and local SDCs that start with
initiators as its structural generators, and above, I have mentioned
other possible structural generators of this kind through the hybrid
EnDQT versions. Let's call the strategy that appeals to i) and ii), assuming
these kinds of structural generators, the local structural generative
strategy since it can allow these theories to respect relativistic
causality. They can also deal in a similar way to EnDQT with the argument from \cite{Maudlin_2014}. See \cite{Pipa2023ATheory} for more details.

On the other hand, WR and PO lack this benefit. If the wave function is a
real field, then non-local causation in spacetime seems built into its
structure. The fact that PO proponents regard the wave function/quantum
states as a law rather than just helping to represent and infer
features of systems and local DSs and ISs (like EnDQT assumes), presses
one to consider that the regularities at a distance in Bell-like
scenarios lead to non-local influences between events.

Thus, via the local structural generative strategy, GQT opens new and
interesting possibilities and strategies compatible with relativistic
causality. Hence, GQT, in principle, provides ways of helping construct
QTs compatible with relativistic causality. Given the
importance of relativistic causality, should be regarded as an important
benefit, and it is one that these views lack.

Third, notice that not all QTs reify the wave function, and adopt WR, such as EnDQT, single-world
relationalist theories mentioned above, such as Relational Quantum
Mechanics, and hybrid classical-quantum theories. Furthermore, not all
QTs see the wave function as a law, such as at least Relational Quantum
Mechanics, other single-world relationalist views, and EnDQT. Since
facts are relative to some entity in single-world relationalist views, is not plausible to view a wave function as governing or describing in
general all those facts.\footnote{See the paragraphs above for the
  reasons why EnDQT doesn't adopt the nomological view.} Furthermore,
unlike the case of the PO, some QTs tend to take systems with
indeterminate values/undifferentiated quantum properties either
absolutely or relationally as playing an important explanatory role,
such as EnDQT or Relational Quantum Mechanics, respectively. Thus, GQT
has the benefit of, in principle, being an ontological framework that
has a wider application, which can facilitate the comparison between QTs
because we can use the same ontological framework to better compare the
different QTs (more on this below).

Fourth, by not reifying wave functions or considering them as laws, and
by allowing for certain new kinds of entities, GQT provides new and
interesting ways of comparing different QTs and finding
their advantages and disadvantages, and which ones we should prefer. I
regard this as another benefit of GQT since it might help us find the best QT, and as we will see, this way of comparing QTs cannot be done
via the other ontological frameworks.

One type of new comparison that GQT allows for is at the level of
generators and generative quantum properties. Despite our world being
fundamentally quantum or at least mostly quantum, certain determinate
values seem to arise, preferably due to certain generators, and
generative quantum properties. What selects the elements of this subset
of determinate values, generators, and generative quantum properties? As
I have argued in Section 2.4, EnDQT via initiators and interactions between systems that belong to SDCs can, in principle,
explain this selection in a unificatory and simple way.\footnote{See
 \cite{Pipa2023ATheory} for more formal details.} It is simple because only one
generator is, in principle, initially and fundamentally postulated (modulo
future developments in cosmology), the inflaton, and simple CDCs. It is
unificatory because all generative quantum properties and generators
trace back to this system, as well as the determinate values that the
systems having them favor.

On the other hand, generative-GRW, generative-Bohmian mechanics, and
generative-hybrid favor only a subset of all quantum properties as
generative, and it is a brute fact why only some quantum properties among
the many existing ones are generative.\footnote{One might argue that there is something special about position, but I am not sure that's right. We can also argue that energy and momentum are also very special.} Also, they postulate many systems
as generators (all the systems with position quantum properties).
Generative-MWI and generative-relationalist-single-world theories don't
postulate fundamental generators, except generative-RQM, which considers
that all systems are generators. However, contrary to EnDQT, they need to provide a special role to many dynamical laws in order to provide a unificatory explanation for why, in a wide range of
interactions, specific systems are generators and others aren't and why
certain quantum properties are generative ones and others aren't. More concretely, what explains these features are the
dynamical laws concerning the quasi-classical domain, which involves Hamiltonians or Lagrangians.
However, this makes these views reliant on many laws to explain these features, which are all the laws that account for decoherence (which includes the various laws that describe emergent features and the diverse terms in the Lagrangians and Hamiltonians). This reliance on many laws might be considered problematic to some who want simpler and more straightforwardly concrete facts to explain generators and generative properties. 

On the other hand, EnDQT explains why certain systems are generators and have generative quantum properties without necessarily relying on many laws. Rather, fundamentally, it appeals to more concrete entities, i.e., quantum systems. Generators and generative quantum properties are explained through chains of interactions that start with the initiators. This general, straightforwardly concrete, and simple facts, seems to me that it renders EnDQT more parsimonious than these views.


A MWI proponent might attempt to
explain generators and generative quantum properties in a more unified way by appealing to the
features of the environments that monitor a target system S.\footnote{For
  example, one could be tempted to adopt the quantum Darwinist strategy
 \cite{Zurek2009QuantumDarwinism} and consider generative quantum properties as those that
  tend to proliferate in an environment.} These environments, via
interactions, select a pointer observable that represents a quantum
property of $S$ that, in later interactions between $S$ and other systems,
will give rise to $S$ having generative quantum property and $S$ being a
generator. However, then one would need to explain why the environment
has those features that give rise to such selection, i.e., those
generative quantum properties and this gives rise to circularity.

One might object that EnDQT may move the brute facts concerning generators
and generative quantum properties to the early universe where initiators
manifested themselves. However, this is at least an explanation for them
(versus a brute fact) or arguably a more parsimonious one since, as I
have mentioned above, fundamentally, we may just need one special generator if we adopt initiators of the kind a) (i.e., the ones that have the DC regarding any system).
Every other fact regarding generators and generative quantum properties
should be explained through chains of interactions that started with the
initiators. Furthermore, any QT already needs to invoke the initial
conditions as a brute fact for various explanatory purposes (e.g.,
explain the arrow of time, solve the problems that inflation pertains to
solve, etc.). EnDQT at least has the advantage of being able to ground
the different already appealed brute facts in a more fundamental one,
which concerns the initial state of initiators. However, note that there
is a sense that this can be an explanation for the initial conditions of
the universe because non-fundamental \emph{special} facts about the
initial conditions of the universe can be grounded on the more
fundamental \emph{special} facts about QT. This is because initiators as
special entities, and the phenomena that they give rise to are
fundamental for QT, according to EnDQT.\footnote{See \cite{Pipa2023ATheory} for more details on these virtues.}

Another type of new comparison that GQT allows for is at the level of
the determination and indetermination structures appealed by each QT.
EnDQT, via initiators and local interactions between systems, explains
the local structure of SDCs in a unified and non-relational way. The ISs
and/or DSs of other QTs, except
generative-single-world-relationalist theories and generative-local-MWI,
have a more complicated non-local structure (e.g., they postulate UDIs,
destruction interactions, etc.), potentially conflicting with
relativity.
Furthermore, generative-single-world-relationalist theories and the
generative-local-MWI offer us DSs that don't causally connect
``distantly separated'' systems or worlds (if a notion of distance even
makes sense for these views) in the sense of worlds or systems that
don't share the same environment. This threatens the power of their
explanatory resources since some local phenomena are more plausible
to explain if they are due to some ``distant'' systems or worlds. I
am not just referring to phenomena involved in Bell correlations here, but whatever is
happening with systems that are $not$ connected with some DSs, although
it is plausible and simpler to consider that they end up influencing the
systems that belong to them. For example, it is plausible to consider that the
values of the sun or even the moon are determinate before they
influence systems on Earth. However, for generative-single-world-relationalists (such as RQM), the systems that belong to the sun or moon at least don't immediately interact with the measurement devices or inhabitants of Earth, and thus their values are indeterminate relative to us. This is at odds with the way we seem to successfully causally explain various phenomena via our scientific theories.

In the case of generative-local-MWI, the threat comes from the events
that are happening in some worlds that seem to influence other worlds
(e.g., the different branches concerning whatever is possibly happening
to my family in the other continent that seem to end up influencing my
branches). However, unlike generative-global-MWI, which connects
different systems within a world via DSs, due to the ever-present
unconnected branching, we don't have a way to track or even make sense
of how different branches causally connect. This is again at odds with the way we seem to successfully causally explain various phenomena via our scientific theories.\footnote{A tension is found
  in the MWI theories between allowing for more explanatory power via a
  generative-global-MWI view and allowing for more locality via a
  generative-local-MWI or arguably a generative-quasi-local-MWI view.} We also have seen above the problems with the consciousness causes collapse theories that were proposed here.

Let's consider that if theory T1 is more parsimonious, less problematic,
and more explanatory than theories T2, then theory T1 should be
preferred to T2. Assuming this, I think that if we consider the above
reasons of locality, parsimony, and explanatory power, we should prefer
EnDQT to the above theories. GQT makes those reasons more manifest.\footnote{What
  about if we compare EnDQT with the hybrid views presented above, and whose comparison hasn't been made yet? We
  would have to examine other costs of these theories. In the case of
  EnDQT-MWI, contrary to EnDQT, it will have an additional problem of
  probabilities plaguing the MWI, which can still make it an undesirable
  view, and it will have the same explanatory power deficits identified
  above that the local-MWI has. Note that due to its determinism and
  reversibility of quantum states, EnDQT-MWI is a theory different from
  EnDQT. In the case of EnDQT-collapse, in so far it would be
  empirically satisfactory, contrary to EnDQT, it will still involve the
  cost of modifying the fundamental equations of QT to account for the
  postulated probability per unit time of a system becoming an
  initiator. Also, it is unclear if this version is really local because
  the probability per unit time of a system becoming an initiator would
  be specified relative to a preferred reference frame. In the case of
  EnDQT-Hybrid, it is still unclear what would constitute its initiators
  or events mentioned above and whether gravity should be treated as a
  separate field that shouldn't be quantized.}

Thus, by not reifying wave functions or considering them as laws and by
allowing for new entities, GQT provides new and interesting ways of
comparing different QTs, finding their advantages and
disadvantages, and deciding which ones to prefer. I regard this as
another benefit of GQT since it might help us find the correct QT. Relatedly, it provides arguments in favor of EnDQT.

However, we might not take the above comparison seriously and object
that the preference for EnDQT regarding the above features when we
compare it with other QTs, disappears when we adopt an ontology that
views the wave function like WR or PO. These QTs can postulate the
existence of the wave function of the universe either as a fundamental
law or field, which provides a simpler and unificatory explanation for
why certain systems are generators and others not, why certain quantum
properties are generative and others aren't, and why certain structures
exist. On top of that, one may even dismiss GQT because it is an
ontological framework that gives, in a sense, an uncharitable treatment
of some QTs that were built under the assumption that we should reify
the wave function in some sense. However, I think the above comparison
should be taken seriously, as well as GQT as a good ontological
framework for QT.

First, I think that multiple issues regarding the different QTs manifest
in similar ways when they adopt WR or the PO. The appeal to brute facts
about primitive ontologies or the reliance on many laws are
still there when these QTs adopt these ontological frameworks. The non-locality at the level of spacetime too, as well as the above issues with single-world-relationalist theories and the local-MWI. GQT just makes these features more manifest.

Furthermore, as I also mentioned, GQT, in principle, facilitates and
improves the comparison between QTs because it can be applied more
widely while using the same kind of ontology. So, it is plausible to
consider that GQT can provide a more charitable treatment of QTs in
general than these other two frameworks. Furthermore, given the current
state of the foundations of physics, where we are still trying to unify
QT with general relativity and solve the measurement problem, I think
that we should consider that the general applicability and comparison
between QTs that GQT allows for is epistemically more valuable than the
restricted applicability that PO and WR tend to lead to. This is because
it might allow for progress by offering new means to evaluate in general
different QTs.

Even if one insists on the simplicity of the wave function/quantum
states, these objects aren't necessarily simpler than DSs and ISs by
many measures of simplicity. For instance, the relations of influence
that they permit, which these ontologies take seriously, are very
diverse and subject to multiple precisifications, not necessarily
simplifying them. Just look at the different QTs, as well as subversions
of them. For instance, just within the MWI, \cite{Sebens2018Self-locatingMechanics}
make the distinction between local and global branching as one way of
precisifying what these objects represent and the relations of influence
that they allow for. There are likely various other ways of precisifying
the relations permitted by the wave function just within the MWI. Also,
these objects per se permit complicated nomic relations, actions at a
distance, and/or evolutions within spaces of many dimensions.
Furthermore, since we have a redundancy in the global phase,\footnote{As
  said in Section 2.1, given the inferential role of quantum states assumed
  by GQT, we can ignore the global phases of quantum states to make
  inferences about and represent properties of systems.} many different
wave functions/quantum states seem to be able to give rise
to/govern/describe the \emph{same} physical state of the systems, and so
in this aspect, these ontologies seem to complexify even more the
description of these relations of influence because there are too many
possibilities. So, given the above reasons, DSs and ISs don't
necessarily give rise to more complicated relations of influence than
the wave function/quantum states.

On top of this complexity, PO and WR have epistemic issues that GQT
doesn't have. Let's suppose we attempt to answer the question regarding
how we know and why we have \emph{this} wave function of the universe and
not another, where this wave function accounts for the behavior of
quantum systems. The answer to this question will hardly be satisfactory
because wave functions aren't directly observable and do not easily
connect with our familiar world or nomic standards. On the other hand,
GQT appeals to, in principle, at least more familiar or standard
entities: systems and their interactions; and a more familiar and
standard view of quantum states to physicists, i.e., mostly as
predictors and inferential tools. Also, many physicists are used to
thinking about indeterminacy via the EEL (Section 1). These standard and
familiar assumptions will likely lead GQT to be considered overall more
satisfactory than PO and WR.

Furthermore, as I have mentioned, unlike GQT, these ontologies give rise to
potentially problematic unexplained non-local influences (PO with its
nomic view),\footnote{Whereas WR and GQT appeal to certain entities.} a
revisionist attitude towards laws (PO), or the problem of making sense
of the three-dimensional space manifest image (WR), which gives rise to
further issues when comparing the different views. So, given this reason
and the ones explained in the two previous paragraphs, I think that the
above unificatory explanation based on the wave function is more
problematic and not necessarily simpler.

Thus, since GQT provides benefits that these influential
frameworks don't offer, including having wider applicability without the
costs mentioned above, which include getting rid of what can be seen as
problematic distractions associated with reifying the wave function or
viewing it as nomic, I think it is a good ontology to compare different
QTs and for QT. Moreover, I also think that GQT and the above comparison
between QTs should be taken seriously.

\section{Conclusion and future
directions}\label{conclusion-and-future-directions}

I have presented Generative Quantum Theory as a new ontology for quantum
theories and have shown how it can be implemented via GRW, the MWI and
single-world relationalist views, Bohmian Mechanics, hybrid
classical-quantum theories, and EnDQT. I have also distinguished it from
the most discussed ontologies for QTs, namely, wave function
realism and primitive ontology, and argued that it has certain benefits
that they lack without some of their costs, such as non-locality.







I have presented generative quantum theories that adopt a
property ontology based on quantum properties, which also allowed for a
new analysis of quantum indeterminacy and determinacy. However, other
generative theories are possible for other kinds of property ontologies.
Future work should explore whether it is beneficial to build generative
theories that use another account of properties, such as
determinable-determinates,\footnote{In this property ontology, we would
  view observables as representing determinables (e.g., position,
  momentum, energy, etc.) and determinate values as representing
  determinates of those determinables. Interactions give rise to a
  determinable with a determinate. Systems would be considered as
  collections of determinables, which at different moments of time, have
  determinates or not (e.g., having a spin-x with or without a
  determinate of spin-x) depending on their interactions like in the
  gappy version of quantum indeterminacy presented in \cite{Calosi2019QuantumIndeterminacy}. Quantum indeterminacy arises when we have state of affairs
  constituted by a system lacking a determinate of a determinable.} etc.

Also, it should explore applying this framework to other QTs, such as
superdeterministic, retrocausal,\footnote{See, e.g., \cite{Hossenfelder2020RethinkingSuperdeterminism, sep-qm-retrocausality}.} and other relationalist
theories, as well as to different pictures of QT. Relatedly, it should
analyze other possible generators (structural and non-structural) and
initiators. This
could allow us to generate further new QTs. Also, it should
extend this view to Quantum Field Theory.\footnote{However, see Section 2.4.}
Furthermore, it should compare GQT with other ontological frameworks
that weren't discussed here. I suspect that GQT will provide many
benefits that they don't have without their costs, given the
distinctiveness of GQT and since, in different ways, most of the other
ontological frameworks reify the wave function or the quantum state or
see it as a law.

\bibliographystyle{plainnat} 
\bibliography{references} 

\end{document}